\documentclass[aip, amsmath,amssymb,
floatfix,     
nofootinbib,  
reprint,
]{revtex4-1}
\usepackage{amsthm}  

\usepackage[total={6.5in,8.75in}, top=1.2in, left=0.9in, includefoot]{geometry}
\usepackage{graphicx}
\usepackage{tikz}
 \usepackage[T1]{fontenc}    

\usepackage{url}
\usepackage[pdftex, bookmarks, colorlinks,pdfencoding=auto, psdextra]{hyperref}  

\usepackage{xcolor}

\newcommand{\Eq}[1]{(\ref{eq:#1})}

\newcommand{\Lem}[1]{Lem.~\ref{lem:#1}}

\newcommand{\Sec}[1]{\S \ref{sec:#1}}
\newcommand{\Fig}[1]{Fig.~\ref{fig:#1}}
\newcommand{\Tbl}[1]{Table~\ref{tbl:#1}}
\newcommand{\App}[1]{Appendix~\ref{app:#1}}

\usepackage{enumitem}
\usepackage{hyperref}
\newcounter{dummy}
\makeatletter 
    \newcommand\myitem[1][]{\item[#1]\refstepcounter{dummy}\def\@currentlabel{#1}}
\makeatother

\newcommand{\InsertFig}[4]
{\begin{figure*}[h!t]
       \centerline{
         \includegraphics[width=#4\textwidth]{./figures/#1}
       }
       \caption{{\footnotesize  #2}
       \label{fig:#3}}
\end{figure*}}

\newcommand{\InsertFigTwo}[5] {
\begin{figure*}[h!t]
       \centerline{
         \includegraphics[width=#5\textwidth]{./figures/#1}
         \hskip 0.2in
         \includegraphics[width=#5\textwidth]{./figures/#2}
       }
       \caption{{\footnotesize  #3}
       \label{fig:#4}}
\end{figure*}}
\newcommand{\InsertFigThree}[6] {
\begin{figure*}[ht]
       \centerline{
         \includegraphics[width=#6\textwidth]{./figures/#1}
         \hskip 0.15in
         \includegraphics[width=#6\textwidth]{./figures/#2}
         \hskip 0.15in
         \includegraphics[width=#6\textwidth]{./figures/#3}
       }
       \caption{{\footnotesize  #4}
       \label{fig:#5}}
\end{figure*}}
\newcommand{\InsertFigFour}[7] {
\begin{figure*}[h!t]
       \centerline{
\renewcommand{\arraystretch}{0.01}
         \begin{tabular}{cc}
         \includegraphics[width=#7\textwidth]{./figures/#1} &  
         \includegraphics[width=#7\textwidth]{./figures/#2} \\
        \includegraphics[width=#7\textwidth]{./figures/#3}  &  
        \includegraphics[width=#7\textwidth]{./figures/#4}
        \end{tabular}
       }
       \caption{{\footnotesize  #5}
       \label{fig:#6}}
\end{figure*}}

\newcommand{\InsertFigSixH}[9] {
\begin{figure*}[ht]
       \centerline{
\renewcommand{\arraystretch}{0.01}
         \begin{tabular}{ccc}
         \includegraphics[width=#9\textwidth]{./figures/#1} &  \includegraphics[width=#9\textwidth]{./figures/#2} &
        \includegraphics[width=#9\textwidth]{./figures/#3}  \\ \includegraphics[width=#9\textwidth]{./figures/#4} &
        \includegraphics[width=#9\textwidth]{./figures/#5}  &  \includegraphics[width=#9\textwidth]{./figures/#6}
        \end{tabular}
       }
       \caption{{\footnotesize  #7}
       \label{fig:#8}}
\end{figure*}}

\newcommand{\InsertFigSixV}[9] {
\begin{figure*}[ht]
       \centerline{
\renewcommand{\arraystretch}{0.01}
         \begin{tabular}{ccc}
         \includegraphics[width=#9\textwidth]{./figures/#1} &  \includegraphics[width=#9\textwidth]{./figures/#2} \\
        \includegraphics[width=#9\textwidth]{./figures/#3}  & \includegraphics[width=#9\textwidth]{./figures/#4} \\
        \includegraphics[width=#9\textwidth]{./figures/#5}  &  \includegraphics[width=#9\textwidth]{./figures/#6}
        \end{tabular}
       }
       \caption{{\footnotesize  #7}
       \label{fig:#8}}
\end{figure*}}

\newcommand{\bR}{{\mathbb{ R}}}

\newcommand{\cO}{{\cal O}}

\newcommand{\hen} {H\'enon }

\newcommand{\jac}{{d}}

\DeclareMathOperator{\Tr}{\mbox{tr}}

\newtheorem{thm}{Theorem}
\newtheorem{lem}[thm]{Lemma}

\newcommand{\citeInline}[1]{Ref.~[\onlinecite{#1}]}
\newcommand{\citesInline}[1]{Refs.~[\onlinecite{#1}]}

\newcommand{\beq}[1]{\begin{equation}\label{eq:#1}}
\newcommand{\eeq}{\end{equation}}

\newenvironment{se}[1]{\equation\label{eq:#1}\aligned}{\endaligned\endequation}
\newcommand{\bsplit}[1]{\begin{se}{#1}}
\newcommand{\esplit}{\end{se}}

\newenvironment{example}[1][]
  {
	\setlength \leftmargini {1.0em}		
	\setlength \topsep {0.5em}			
	\begin{quote}
	{\it Example#1} }
	{\end{quote}
  }
\newcommand{\bexam}[1][:]{\begin{example}[#1]}
\newcommand{\eexam}{\end{example}}

\begin{document}

\title{Visualizing Attractors of the Three-Dimensional Generalized H\'enon Map}
\author{Amanda E. Hampton}
\email{	Amanda.Hampton@colorado.edu}
\author{James D.~Meiss}
\email{ James.Meiss@colorado.edu}
\affiliation{ Department of Applied Mathematics\\
    University of Colorado \\
	Boulder, CO 80309-0526}

\date{\today}

\begin{abstract}
We study dynamics of a generic quadratic diffeomorphism, a  3D generalization of the planar \hen map. Focusing on the dissipative, orientation preserving case, we give a comprehensive parameter study of codimension-one and two bifurcations.
Periodic orbits, born at resonant, Neimark-Sacker bifurcations, give rise to Arnold tongues in parameter space. Aperiodic attractors include invariant circles and chaotic orbits; these are distinguished by rotation number and Lyapunov exponents.
Chaotic orbits include H\'enon-like and Lorenz-like attractors, which can arise from period-doubling cascades, and those born from the destruction of invariant circles. The latter lie on paraboloids near the local unstable manifold of a fixed point.
\end{abstract}

\maketitle

\begin{quotation}
Since H\'enon's observation of strange, chaotic attractors in his eponymous  two-dimensional map,\cite{Henon76} this quadratic map has become a pivotal model to understand chaotic dynamics in the plane, see \citesInline{Guckenheimer02, Kuznetsov04}. It is of much interest to understand the dynamics of higher-dimensional generalizations of this map as models for the onset and development of chaotic dynamics. As a prototypical model, we study the so-called 3D generalized \hen map,\cite{Gonchenko05b} which is a quadratic normal form for several bifurcation scenarios. Extending upon previous work, we study the periodic and aperiodic dynamics of this map using a variety of visualizations in both parameter and phase space. For example, Arnold tongues, or resonant regions, in parameter space correspond to attracting periodic orbits in phase space. We compute heteroclinic trajectories between orbits in the tongues as well as in period-doubling cascades. We also follow the evolution of invariant circles as they bifurcate, some of which can become complex chaotic attractors.
\end{quotation}

\section{Introduction}\label{sec:Intro}

The two-dimensional H\'enon map\cite{Henon76} is \textit{the} quadratic diffeomorphism of the plane: every such diffeomorphism is conjugate to a map in the two-parameter H\'enon family. Prominent rigorous results for this diffeomorphism include that of Devaney and Nitecki, \cite{Devaney79} who showed that the dynamics of its bounded orbits are conjugate to a Smale horseshoe in some parameter regimes, and that of Benedicks and Carleson,\cite{Benedicks91} who showed that when the Jacobian is sufficiently small there are cases for which the map has a transitive attractor with a positive Lyapunov exponent. It was also shown that the horseshoe in this map
can be thought of as arising from an anti-integrable limit, where the dynamics becomes
non-deterministic.\cite{Sterling98}

Some of these results have been generalized to higher-dimensional maps. For the three-dimensional (3D) case it was shown in \citeInline{Lomeli98} that every quadratic diffeomorphism with a quadratic inverse is conjugate to the map $L: \bR^3 \to \bR^3$
\bsplit{Q3DMap}
	L(x,y,z) &= (\delta z+ G(x,y), x,y), \\
	G(x,y)   &= \alpha + \tau x - \sigma y + ax^2 + bxy + cy^2 .
\esplit
This analysis was later generalized to include 3D quadratic diffeomorphisms that have quartic inverses, \cite{Elhadj09} and to higher dimensions. \cite{Lenz99}
The diffeomorphism \Eq{Q3DMap} arises as a normal from in the volume preserving ($\delta = 1$) case near a fixed point with three unit multipliers.\cite{Dullin08a,Dullin09a} It has also been shown to be a  normal form near a map with a saddle-focus fixed point that has a quadratic homoclinic tangency. \cite{Gonchenko05b,Gonchenko06}
The theory of anti-integrability also applies to these maps when
$\alpha \to -\infty$, see e.g., \citeInline{Hampton22}.

The map $L$ has seven parameters, six in the quadratic polynomial $G$ and $\delta = \det{DL}$, the Jacobian determinant. This parameter set can be reduced, as noted in \citeInline{Lomeli98}: whenever $a+b+c\neq 0$ and $2a+b\neq0$,\footnote
{If one of these is violated, other scaling transformations can be found to eliminate two of the parameters.}
an affine coordinate transformation allows one to set 
\beq{ParamSpace}
    a+b+c = 1 \mbox{ and } \tau = 0 .
\eeq
Using this simplification, \Eq{Q3DMap} depends only on $(\alpha, \sigma, a, c)$ and the Jacobian $\delta$.

The case $(a,b,c) = (1,0,0)$ has been called ``the 3D H\'enon map'' since it only has one nonlinear term, similar to the classical H\'enon map.\cite{Henon76}
In a series of papers starting with \citesInline{Gonchenko05b, Gonchenko06}, Gonchenko and collaborators focus on the formation of chaotic attractors for this map.\footnote
{
    In their notation $M_1 = -\alpha, B = \delta, M_2 = -\sigma$.
}
These attractors are shown to be ``wild hyperbolic''\footnote
{Nearby maps have Newhouse tangencies between its stable and unstable manifolds. \cite{Turaev98}} 
near the parameters $(\alpha,\sigma,\delta) = (\tfrac14,-1,1)$, where a pair of fixed points are born with multipliers $(-1,-1,1)$. These attractors are also ``pseudo-hyperbolic,''\footnote
{Its tangent space splits into a a strong stable subspace and a complementary subspace that exponentially expands volume.}
like the Lorenz attractor, implying that every orbit in the attractor has at least one positive and one negative Lyapunov exponent. According to \citeInline{Gonchenko16}, there are five types of such attractors formed from the unstable manifold of a fixed point, including ``Lorenz-like'' and several ``figure-eight'' attractors.\cite{Gonchenko13b} Moreover, there are infinite cascades in parameter space of nearby systems with Lorenz-like attractors.\cite{Gonchenko17} Chaotic attractors for the  non-orientable case have also been studied.\cite{Gonchenko21d}

Following Gonchenko et al., we will primarily study the 3D H\'enon case using the scaling \Eq{ParamSpace} and taking $0 < \delta < 1$, so that the map is volume contracting and orientation preserving.
We focus on two examples:
\begin{enumerate}\label{StdParamters}
    \myitem[(SC)]\label{SmallDeltaCase} Strongly Contracting:  $(a,c,\delta)=(1,0,0.05)$,
    \myitem[(MC)]\label{LorenzLikeCase} Moderately Contacting: $(a,c,\delta)=(1,0,0.7)$,
\end{enumerate}
allowing $(\alpha,\sigma)$ to vary.
In \Sec{GeneralBifTheory}, we recall basic bifurcation behavior of periodic orbits for a 3D map using the trace and second trace of the Jacobian. These bifurcation conditions are transformed to conditions on $(\alpha,\sigma)$ for the fixed points of \Eq{Q3DMap} in \Sec{BifofFixedPts}.
Similar bifurcation criterion were obtained for the 3D H\'enon map in \citeInline{Zhao17} for the case that $\sigma = \delta$, as well as in \citeInline{Richter02} when $\sigma=0$ for three and more dimensions.
In \Sec{BddOrbits}, we prove that all bounded orbits of the 3D H\'enon map lie within a finite cube about origin, following the proof of a related theorem in \citeInline{Lomeli98}.

The simplest bounded orbits are stable and periodic. Parameter regions in the $(\alpha,\sigma)$-plane containing such attractors, analogous to Arnold tongues, are computed in \Sec{PeriodicAttractors}, where we also compute heteroclinic manifolds between stable and unstable orbits. 
In \Sec{NonresonantAttractors}, we study aperiodic attractors, computing the maximal Lyapunov exponent to distinguish between regular and chaotic cases. Regular aperiodic attractors come in the form of invariant circles, which we study in \Sec{Nonchaotic} extending the volume-preserving case studied in \citeInline{Dullin09a}.
In \Sec{ChaoticAttractors}, we find H\'enon-like attractors for case~\ref{SmallDeltaCase} and  discrete, Lorenz-like attractors\cite{Gonchenko21a} for case~\ref{LorenzLikeCase}. Additionally, we observe chaotic attractors arising from bifurcations of invariant circles that are unlike those in \citesInline{Gonchenko14, Gonchenko21c}; these also do not seem to be related to the generalizations of Smale's horseshoe to 3D found by \citeInline{Zhang16}. 

\section{Bifurcations for 3D Maps}\label{sec:GeneralBifTheory}
For a fixed point $\xi^* = f(\xi^*)$ of a 3D map $f$, the eigenvalues of the Jacobian $A = Df(\xi^*)$ are given
by the characteristic polynomial
\beq{CharPoly}
    p_A(\lambda) = \det(\lambda I - A) = \lambda^3-t\lambda^2+s\lambda-\jac .
\eeq
We refer to these as the \textit{multipliers} of the fixed point.
Here $t = \Tr(A)$ is the trace and $\jac = \det(A)$. An expression for the ``second trace'', $s$, can be obtained from the Cayley-Hamilton theorem: a matrix satisfies its own characteristic polynomial,
$
    A^3-tA^2+sA-\jac I=0
$.
Multiplying this by $A^{-3}$ 
implies that $s = \jac \Tr(A^{-1})$. Finally, multiplying the Cayley polynomial by $A^{-1}$ and taking the trace gives
\beq{tsDfns}
	s  = \tfrac12 \left(t^2-\Tr(A^2)\right) .
\eeq
Of course, the multipliers can be related to the coefficients of \Eq{CharPoly} by the symmetric polynomials,
\bsplit{st-EV}
   t &= \lambda_1+\lambda_2 + \lambda_3, \\
   s &=  \lambda_1\lambda_2 + \lambda_1\lambda_3 + \lambda_2\lambda_3, \\
   d &= \lambda_1\lambda_2\lambda_3 .
\esplit

More generally, for an orbit, $\xi_{t} = f(\xi_{t-1})$, that has period $n$, $\xi_0 = f^n(\xi_0)$,
the Jacobian becomes
\[
   A =  Df^n(\xi_0) = Df(\xi_{n-1})Df(\xi_{n-2})\ldots Df(\xi_1)Df(\xi_0) .
\]
Thus, using the same process as above, we can find a general expression for the trace and second trace of a period-$n$ orbit:
\[
	t = \text{Tr}(Df^n), \quad \quad
	s = \frac{1}{2} (t^2-\text{Tr}((Df^n)^2)) .  
\]


Following \citeInline{Kuznetsov19}, the simplest, local codimension-one bifurcations---saddle-node, period-doubling and Neimark-Sacker---occur when at least one multiplier has unit modulus. Each of these occurs on a surface in $(t,s,\jac)$, given in \Tbl{Bifurcations}. Sections through these surfaces for four values of $\jac$ are shown in \Fig{TSPlanes}; similar figures can be found in \citeInline{Lomeli98} for the case $\jac = 1$ and in \citeInline{Gonchenko16}, where they are referred to as ``saddle-charts.''

\renewcommand{\arraystretch}{1.2}
\begin{table*}[ht]
\centering
\begin{tabular}{l|cc|c|c}

    CoD & \multicolumn{2}{|c|}{Bifurcation} &  Multiplier & $(t,s)$ \\
    \hline

    1 & Saddle-Node &(SN) &  $\lambda_1=1$ & $(t,t+\jac-1)$  \\
     & Period Doubling &(PD) & $\lambda_1=-1$ & $(t,-t-\jac-1)$ \\
      & Neimark-Sacker &(NS) &  $\lambda_{1,2}=e^{\pm 2\pi i \omega}$ 
                                    & $(\jac+2\cos(2\pi\omega),2\jac\cos(2\pi\omega) + 1)$\\
    \hline
    2 & $\omega = \tfrac01$ & (R1) & $\lambda_{1,2}=1$ & $(\jac+2,2\jac+1)$ \\
    
      & $\omega = \tfrac12$ & (R2) & $\lambda_{1,2}=-1$ & $(\jac-2,1-2\jac)$ \\
    
      & $\omega = \tfrac13$ & (R3) & $\lambda_{1,2}=e^{\pm 2\pi i/3}$  & $(\jac-1,1-\jac)$  \\
      & $\omega = \tfrac14$ &(R4) & $\lambda_{1,2}=\pm i $  & $(\jac,1)$ \\
     & Saddle-Node Flip &(SNf) &  $\lambda_1=1,\lambda_2=-1$ & $(-\jac,-1)$ \\
\end{tabular}
\caption{Local bifurcations for a 3D map with Jacobian $\jac$, trace $t$, and second trace $s$. For each $\jac$, codimension-one bifurcations lie on lines in the $(t,s)$ plane, and codimension-two bifurcations at points. Since $\lambda_1\lambda_2\lambda_3=\jac$, for the NS and resonant bifurcations, $\lambda_3=\jac$, and for SNf, $\lambda_3=-\jac$.}
\label{tbl:Bifurcations}
\end{table*}

Saddle-node (SN) bifurcations require a unit multiplier, so that \Eq{CharPoly} gives $p_A(1) = 1-t+s-\jac = 0$. This corresponds to a line in the $(t,s)$-plane, shown in blue in \Fig{TSPlanes}. A period-doubling (PD) bifurcation occurs when there is a $-1$ multiplier, or $p_A(-1) = -1-t-s-\jac=0$; shown in red in \Fig{TSPlanes}. 
The third codimension-one bifurcation, the Neimark-Sacker (NS), occurs when there is a pair of complex multipliers with unit modulus; it generically results in the creation of an invariant circle or pair of periodic orbits with differing stabilities. Since $|\lambda_{1,2}| = 1$,  \Eq{st-EV} gives $\lambda_3 = \jac$, so that $p_A(\jac) = \jac(\jac^2-t\jac+s-1)=0$. The resulting line, $s = \jac(t-\jac) + 1$, is shown in black in \Fig{TSPlanes}.
Solving for the complex multipliers gives
\beq{NSEvals}
    \lambda_{1,2}= e^{\pm 2\pi i\omega}
                 =\tfrac12 \left(t-\jac\pm\sqrt{(t-\jac)^2-4}\right),
\eeq
or that  $t-d = 2 \cos(2\pi\omega)$, for rotation number $\omega$. 

Some codimension-two bifurcations occur along the NS line when $\omega = \tfrac{p}{q}$,
and generically result in the birth of a pair of period-$q$ orbits.
The endpoints of the NS line occur where it intersects the SN and PD lines at $\omega = 0$ and $\omega = \tfrac12$, 
labeled (R1) and (R2), respectively, in \Tbl{Bifurcations}.
The table also shows the period-three (R3) and period-four (R4) cases. 
Finally, the saddle-node flip (SNf) bifurcation point, with multipliers $(-1,1,-\jac)$ is at the intersection of the PD and SN lines.



A curve in parameter space along which there is a double multiplier, say $\lambda_1=\lambda_2=r$, for $r\in\bR$, corresponds to the transition from real to complex multipliers.
Using \Eq{st-EV}, this occurs when $\lambda_3 = d/r^2$ along the parametric curve
\beq{DoubleEval}
    (t,s) =\left(\frac{\jac}{r^2} + 2r ,  
            2\frac{\jac}{r} + r^2\right) .
\eeq
This curve has two branches when $\jac \neq 0$, one of which has a cusp at
$r^3 = \jac$, i.e., at $(t,s) = 3(\jac^{1/3},\jac^{2/3})$.
When $r=1$, this curve is tangent to the SN line and crosses the R1 point, and when $r=-1$ the curve is tangent to the PD line and crosses the R2 point.
Four segments of these curves, labelled by ranges of the double multiplier, are shown in \Fig{TSPlanes}.

We also include representative configurations of the multipliers in the complex plane relative to the unit circle for each region of the $(t,s)$ plane in \Fig{TSPlanes}.  
For $0 \leq d < 1$, there exists a small triangular region with all stable multipliers ($|\lambda_i| < 1$) that is bounded by the SN, PD and NS lines with vertices R1, R2, and SNf. Otherwise, the SN and PD lines divide the plane into four regions where stability types of the multipliers of the fixed points alternate between 
having one unstable and two stable multipliers and two unstable and one stable multiplier.

\InsertFigFour{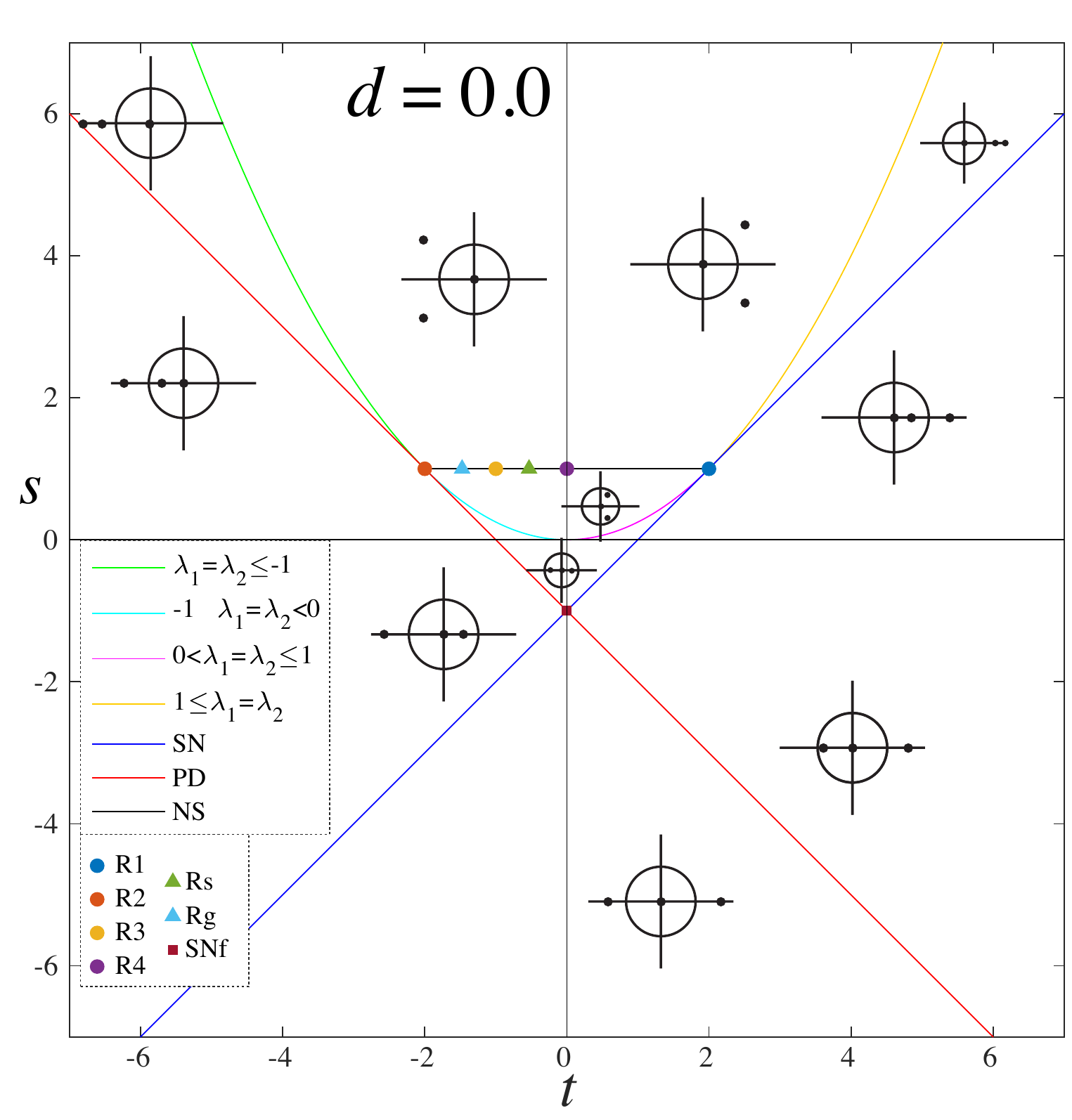}{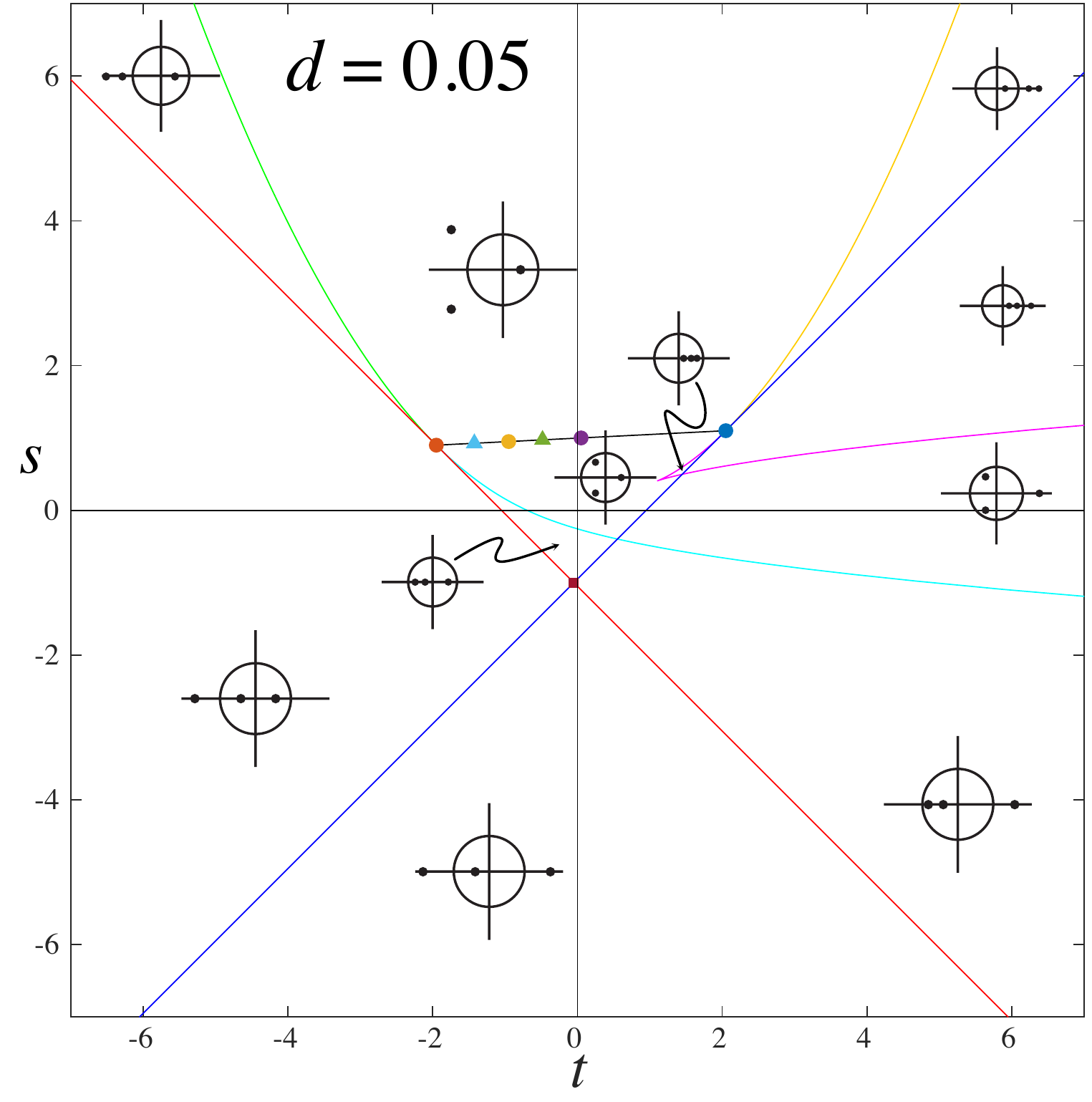}{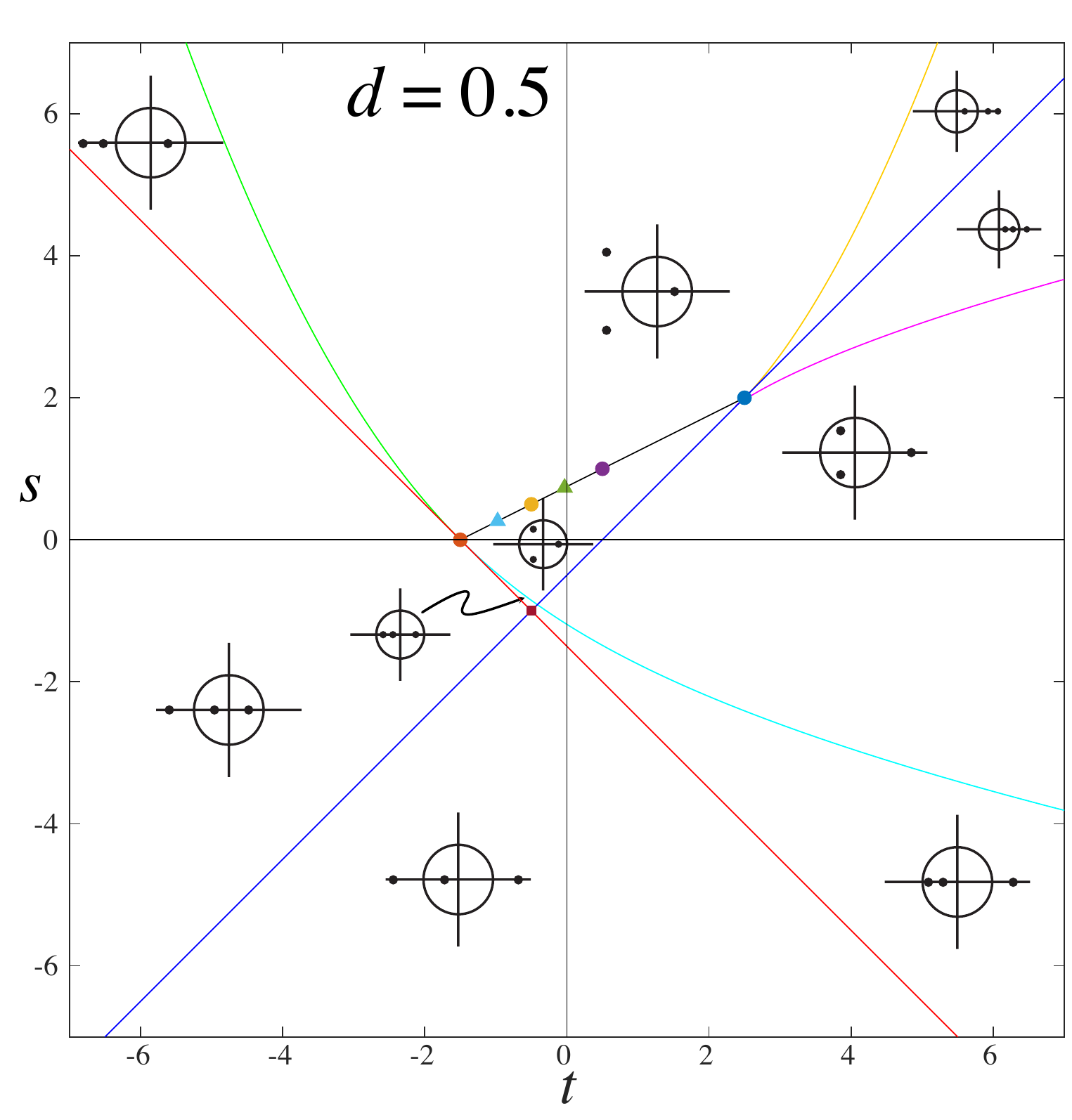}{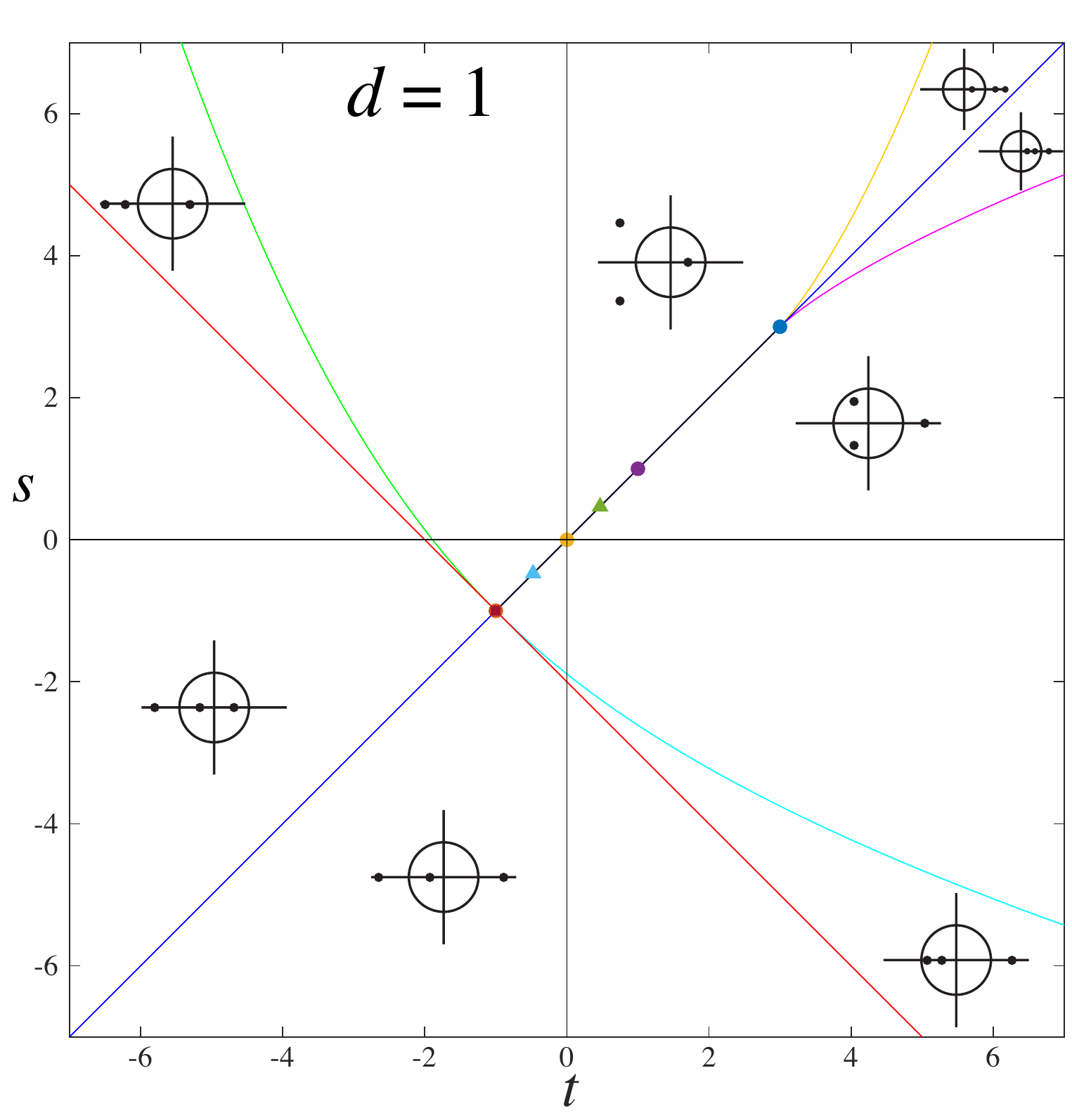}
{Codimension-one and two bifurcations in the $(t,s)$-plane for $\jac=0$, $\jac=0.05$, $\jac=0.5$ and $\jac=1$. The double multiplier curves \Eq{DoubleEval} and irrational NS points for $\omega=(\sqrt{5}-1)/2$ (Rg) and $\omega=1/\sqrt{2}$ (Rs) for the golden and silver means, respectively, are also shown. Behavior at irrational points along the NS line will be discussed in \Sec{NonresonantAttractors}.
The insets show the complex plane with representative multiplier locations relative to the unit circle.}
{TSPlanes}{0.35}

\section{Fixed Points of the Quadratic Map}\label{sec:BifofFixedPts}

We now apply the results of \Sec{GeneralBifTheory} to fixed points of \Eq{Q3DMap} with parameter convention \Eq{ParamSpace}. 
The fixed points have the form $\xi_\pm = (x_{\pm},x_{\pm},x_{\pm})$ where
\bsplit{FixedPoints}
    x_\pm &= \tfrac12 \left(\sigma-\delta+1\right) \pm \sqrt{\alpha_{SN}-\alpha}, \\
    \alpha_{SN} &\equiv \frac{1}{4}(\sigma-\delta+1)^2,
\esplit
provided $\alpha \le \alpha_{SN}$. These are born in a saddle-node bifurcation when $\alpha = \alpha_{SN}$ with $x_{\pm}^2 = x_{SN}^2 = \alpha_{SN}$. 
Thinking of $\alpha$ as a bifurcation parameter, the form \Eq{FixedPoints} implies that there is a fixed point at $x^*$, say, when 
$\alpha = \alpha^* \le \alpha_{SN}$ whenever $(x^*-x_{SN})^2 = \alpha_{SN} - \alpha^*$, or equivalently when
\beq{alphaBif}
    \alpha^* = x^*(2x_{SN}-x^*).
\eeq

The stability of the fixed points is determined by the linearization of \Eq{Q3DMap},
\[
DL=
\begin{pmatrix}
    2ax+by & -\sigma+bx+2cy & \delta \\
    1 & 0 & 0 \\
    0 & 1 & 0
\end{pmatrix} ,
\]
This is in companion form, implying that at a fixed point the trace and second trace are
\beq{tsFixedPts}
    t_\pm  = (2a+b)x_{\pm}, \quad s_\pm  =  \sigma-(b+2c) x_{\pm} ,
\eeq
and the Jacobian is $\jac = \delta$.

The fixed points lie on a line in the $(t,s)$-plane that can be found by eliminating $x_{\pm}$ and $b$ from \Eq{tsFixedPts},
\beq{FixedPttstrace}
    (1+a-c)(s_\pm - \sigma) = -(1-a+c) t_\pm ,
\eeq
and are born at the intersection of this line with the SN line from \Tbl{Bifurcations},
\[
    t_{SN} = \tfrac12 (\sigma-\delta+1)(1+a-c).
\]
Moreover, \Eq{tsFixedPts} gives
\[
    s_\pm-t_\pm- \delta+ 1 = \sigma -\delta +1 - 2x_\pm = \mp 2\sqrt{\alpha_{SN}-\alpha}.
\]
Since $x_-<x_{SN}<x_+$, this implies that $x_+$ lies below and $x_-$ above the SN line in the $(t,s)$ plane. Therefore when $0\leq\delta<1$, if there exists an attracting fixed point, it must be $\xi_-$.

Using the equations for the SN and PD lines in \Tbl{Bifurcations}, the fixed points are born above the PD line when $t_{SN}>-\delta$, or
\[
   (\delta-\sigma-1)(1+a-c)< 2\delta.
\]

Provided $a\neq c$, one of the fixed points undergoes a period-doubling bifurcation at 
\beq{x_pd}
    x_{PD}=\frac{\sigma+\delta+1}{2(c-a)}.
\eeq
An expression for the period doubling value of $\alpha$ is obtained from \Eq{alphaBif}:
$\alpha_{PD} = x_{PD}(2x_{SN}-x_{PD})$.
Note that for a fixed point to `double,' it needs to exist, thus we must have $\alpha_{PD} \leq \alpha_{SN}$. 
In addition, when $x_{PD} < x_{SN}$, the $\xi_{-}$ fixed point doubles, and when $x_{PD} > x_{SN}$, the $\xi_{+}$ point doubles.

Similarly, using \Eq{tsFixedPts} and the expression in \Tbl{Bifurcations} for the NS bifurcation gives
\beq{x_NS}
    x_{NS}=\frac{\sigma-\delta\tau+\delta^2-1}{\delta(1+a-c)+(1-a+c)},
\eeq
provided the denominator is nonzero.
Of course, this bifurcation only exists if the fixed points intersect the NS line on the interval
$\delta-2<t<\delta+2$.
Again, \Eq{alphaBif} can be used to determine the bifurcation value, $\alpha_{NS}$.

When the parameters $(a,c,\delta)$ are fixed, the generic bifurcation diagrams of \Fig{TSPlanes} can be transformed to corresponding diagrams for a fixed point in the $(\alpha,\sigma)$ plane; details are given in \App{ParameterConversion},
including criteria for codimension-two bifurcations.
The results are shown in \Fig{FixPtBifs} for the fixed point $\xi_-$---as this is the only fixed point that can be attracting---for cases \ref{SmallDeltaCase} and \ref{LorenzLikeCase}.


\InsertFigTwo{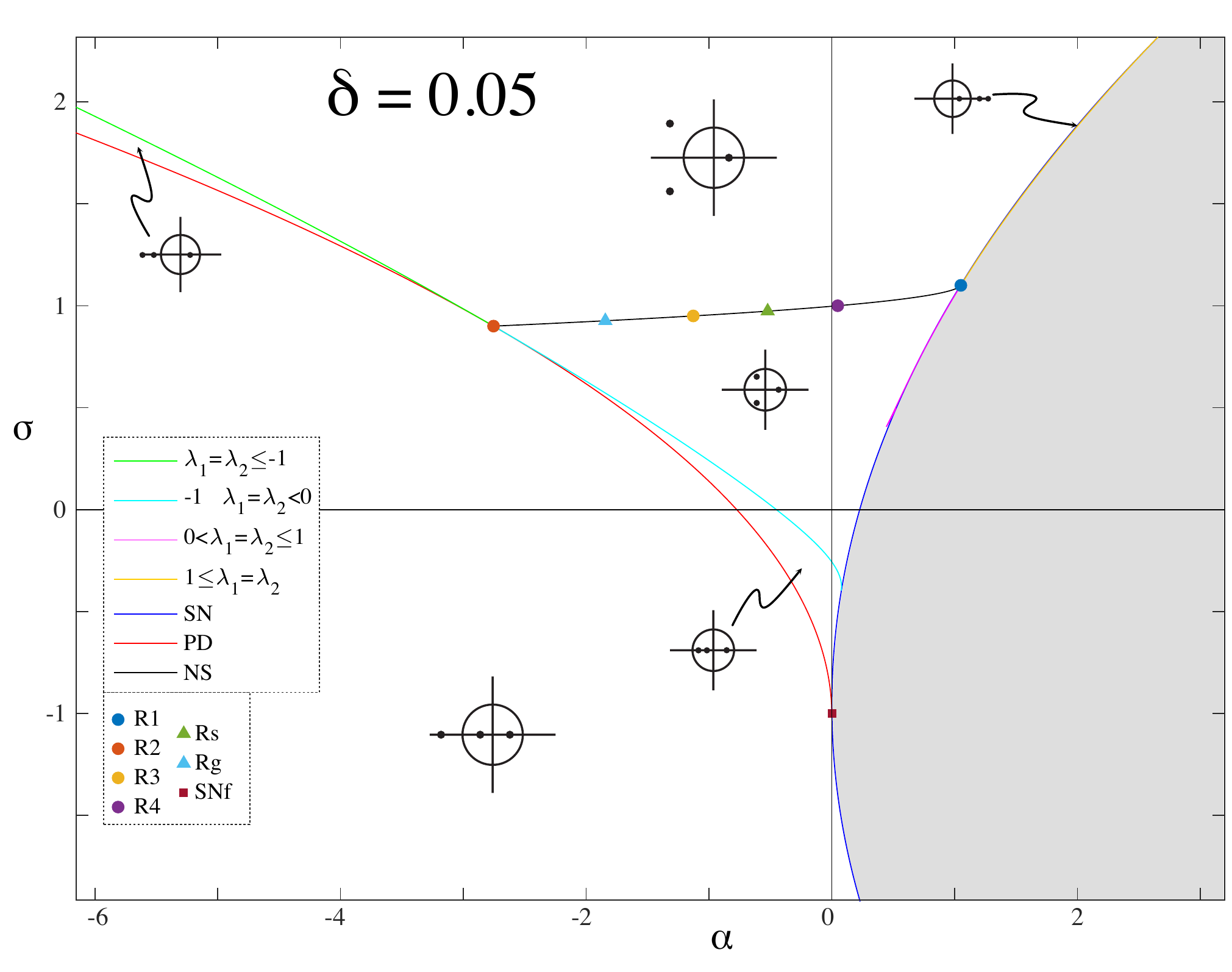}{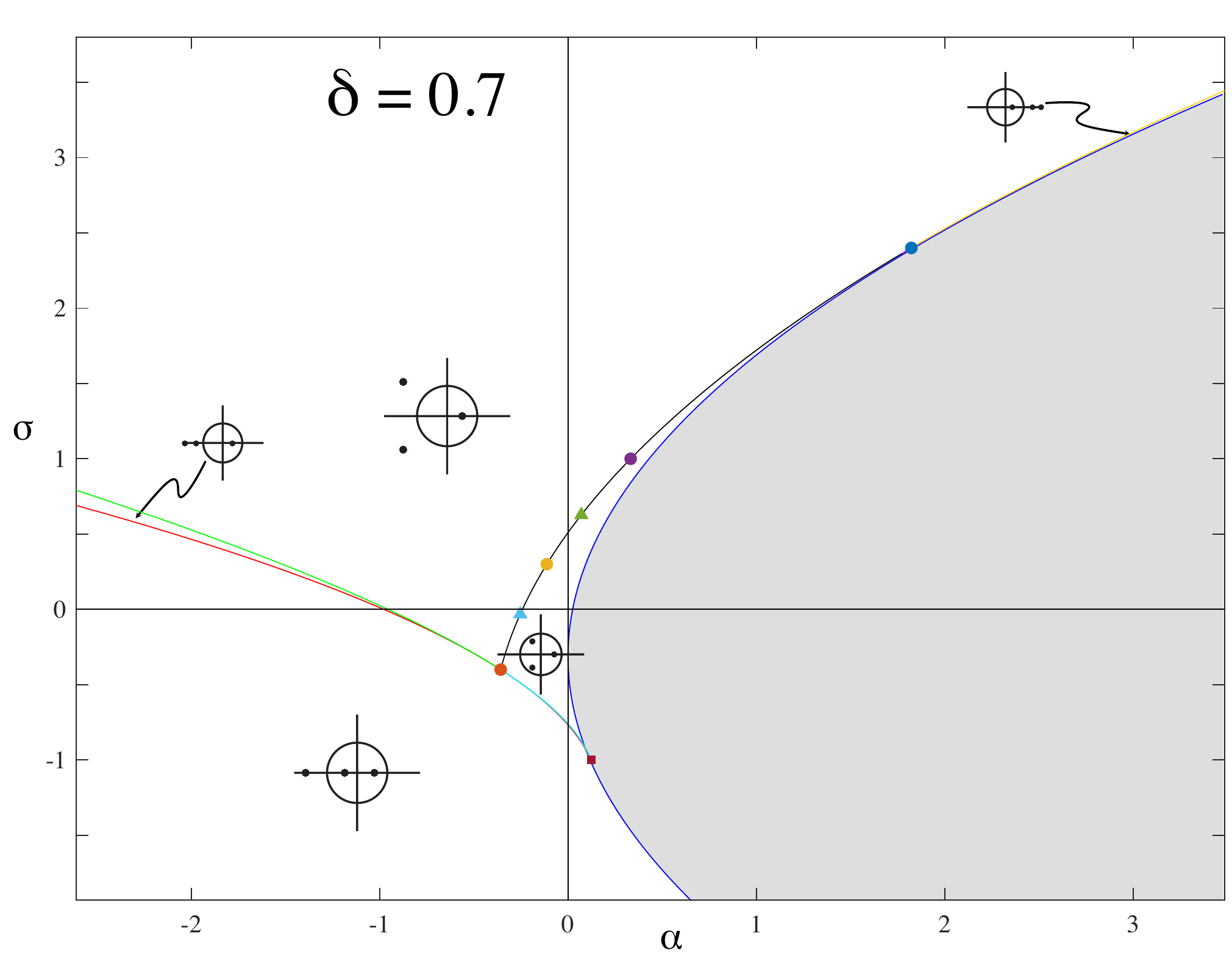}
{Bifurcation curves for the fixed point $\xi_-$ of \Eq{Q3DMap} for the case~\ref{SmallDeltaCase} ($\delta = 0.05$) and \ref{LorenzLikeCase} ($\delta = 0.7$) as $(\alpha, \sigma)$ vary. 
The fixed points do not exist to the right of the SN line, i.e., in the gray region.
The double-multiplier curves \Eq{DoubleEval} are shown for $|r|>\sqrt{\delta}$, as this range corresponds to a double multiplier for $\xi_-$, see \App{ParameterConversion} for details.}{FixPtBifs}{0.5}

\section{Bounded Orbits}\label{sec:BddOrbits}

In \citeInline{Lomeli98}, it was shown that if the quadratic form $Q(x,y)=ax^2+bxy+cy^2$ is positive definite and $\delta = 1$, then
there is a cube that contains all bounded orbits of \Eq{Q3DMap}; equivalently, all points outside of this cube escape to infinity. We show in this section that a similar argument can be used for the case $(a,b,c)=(1,0,0)$, where $Q$ is semi-definite, for any $\delta > 0$.

\begin{lem}\label{lem:BddOrbits} If $(a,b,c)=(1,0,0)$, $\tau = 0$, and $\delta > 0$,
then  all bounded orbits of the map \Eq{Q3DMap} are contained within the cube $\{(x,y,z):\, |x|,|y|,|z| \leq\kappa\}$ where
\beq{kappa}
    \kappa=\tfrac12 \left(|\sigma|+\delta+1 + \sqrt{(|\sigma|+\delta+1)^2+4|\alpha|} \right) .
\eeq
\end{lem}

\begin{proof}
For an orbit $\xi_t = (x_t,y_t,z_t)$ of the map \Eq{Q3DMap}, $y_t = x_{t-1}$ and $z_t = x_{t-2}$, so the map
is equivalent to the forward and backward difference equations
\begin{align}
    x_{t+1} &= \delta x_{t-2} + \alpha  -\sigma x_{t-1} +x_t^2 , \label{eq:ForwardMap} \\
    x_{t-3} &=\delta^{-1}(x_{t} - \alpha + \sigma x_{t-2} - x_{t-1}^2).\label{eq:BackwardMap}
\end{align}
There are three cases to consider that depend on which term in the sequence $x_{t-2}, x_{t-1},x_t$ is largest.
\begin{enumerate}
    \item $|x_t|\geq \max{(|x_{t-1}|,|x_{t-2}|)}$: then \Eq{ForwardMap} gives
    \[
    	x_{t+1} \geq x_t^2-(|\sigma|+|\delta|)|x_t|-|\alpha| ,
    \]
    giving a lower bound to the forward iterate.
    Notice this inequality implies that the next iterate lies above an even, piecewise-parabolic curve with negative vertical intercept. Thus there exists $\kappa>0$ such that whenever $|x_t|>\kappa$, we have
    \[
    	x_{t+1} \geq x_t^2-(|\sigma|+|\delta|)|x_t|-|\alpha| > |x_t|>\kappa .
    \]
    Here $\kappa$ is the maximum root of $x_t^2-(|\sigma|+|\delta|+1)|x_t|-|\alpha|$, given by \Eq{kappa}.
    This implies that $x_{t+1}>|x_t|\geq|x_{t-1}|$. Recursively applying this argument to each forward step implies that
    \[
    	x_{t+k}>x_{t+k-1}> \ldots >x_{t+1}>|x_t|>\kappa .
    \]
   This is a monotone increasing sequence that cannot have a finite limit: if it were to converge, it would have to converge to one of the fixed points $x_\pm$, \Eq{FixedPoints}, but this is impossible since, by simple calculation, $\kappa>|x_\pm|$.
    
    \item$|x_{t-2}|\geq \max{(|x_{t}|,|x_{t-1}|)}$: then \Eq{BackwardMap} gives
    \[
    	x_{t-3}\leq \delta^{-1}(-x_{t-2}^2  + (1 + |\sigma|) |x_{t-2}| + |\alpha|) ,
    \]
    giving an upper bound to the preimage.
    This inequality is the space below an even, downward facing piecewise parabola with a positive vertical-intercept. As before, when $|x_{t-2}|>\kappa$,
    \begin{align*}
        x_{t-3} \leq \delta^{-1}(-x_{t-2}^2  &+ (1 + |\sigma|) |x_{t-2}| + |\alpha|)  \\
                &< -|x_{t-2}| < -\kappa .
    \end{align*}
    Applying this argument recursively implies that $x_{t-k}$ is a monotone decreasing sequence. Again, the sequence must be unbounded.
    
    \item $|x_{t-1}|\geq \max{(|x_{t}|,|x_{t-2}|)}$: using \Eq{BackwardMap} gives
    \[
    	x_{t-3}\leq \delta^{-1}(-x_{t-1}^2  + (1 + |\sigma|) |x_{t-1}| + |\alpha|) < -|x_{t-1}|
    \]
    whenever $-|x_{t-1}|<-\kappa$. Recursively, this implies $x_{t-3}<-|x_{t-1}|\leq-|x_{t-2}|$, resulting in the same scenario as (2).
\end{enumerate}
Thus, bounded orbits of the map \Eq{Q3DMap} with $(a,b,c)=(1,0,0), \tau=0$, and $\delta>0$ must lie within a $\kappa$-cube about origin.
\end{proof}

To illustrate this result, we compute the volume of bounded orbits, i.e. the volume of the basin of any attractors, for the cases \ref{SmallDeltaCase} and \ref{LorenzLikeCase}, see \Fig{BddVolumes}.
For each point on a $500^2$ grid in the $(\alpha,\sigma)$-plane for which the fixed points \Eq{FixedPoints} exist ($\alpha \le \alpha_{SN}$), we choose $50^3$ initial conditions on a uniform grid in a cube with bounds \Eq{kappa} that vary with $(\alpha, \sigma)$.
An orbit is declared to be unbounded if it leaves the $\kappa$-cube within $200$ iterates.

The selected parameters in \Fig{BddVolumes} focus on the triangular regions of \Fig{FixPtBifs} where $\xi_-$ is an attracting fixed point. The bifurcation curves (dashed) and codimension-two points are also shown (refer to the legend in \Fig{FixPtBifs}). 
Note that near the SN line, the volume of bounded orbits in the $\kappa$-cube is small since most are found near the fixed points, which are close together.

\InsertFig{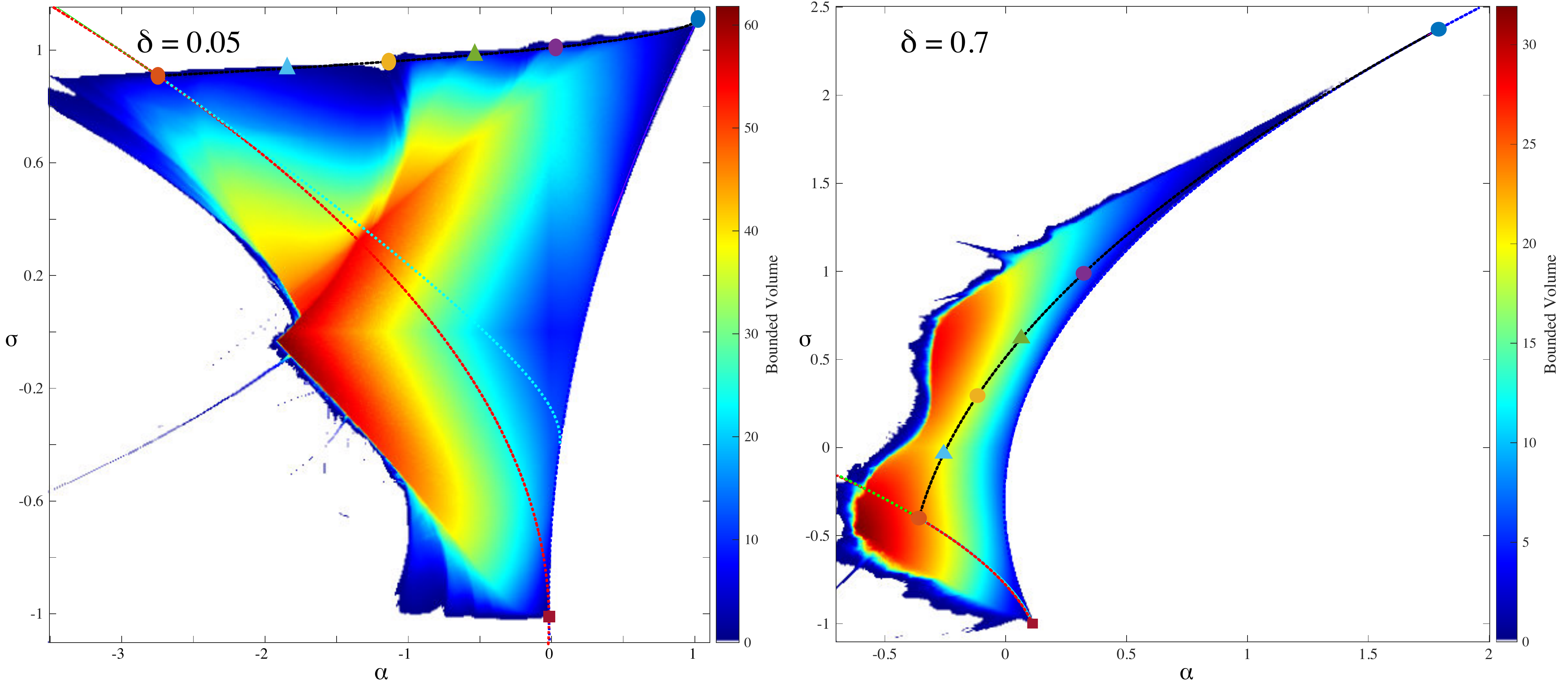}{ The volume of bounded orbits for $50^3$ initial conditions in the $\kappa$-cube for case~\ref{SmallDeltaCase} ($\delta=0.05$) and case~\ref{LorenzLikeCase} ($\delta=0.7$) as a function of $(\alpha,\sigma)$. Bifurcation curves and points from \Fig{FixPtBifs} are also shown. Notice the color bar for case~\ref{SmallDeltaCase} includes larger values, with a maximum volume of $61.8$. This is expected as it is more strongly contracting than case \ref{LorenzLikeCase}, which has a maximum volume of $31.9$.}
{BddVolumes}{0.95}

\section{Periodic Attractors}\label{sec:PeriodicAttractors}

In this section we compute regions in parameter space for which \Eq{Q3DMap} has attracting periodic orbits. If there is a unique, bounded attractor, it can easily be found by choosing an appropriate initial condition within the cube of \Lem{BddOrbits} and iterating until the orbit limits on the attractor. Once the transient is removed, parameter regions of periodic behavior can be computed by looking for recurrence.

Computed periodic regions are shown in \Fig{ResonantEx} for the two cases, \ref{SmallDeltaCase} and \ref{LorenzLikeCase}. To compute these, the initial point is set to $\xi_0 = 0$ when neither of the fixed points exist (i.e., to the right of the SN line in the  $(\alpha, \sigma)$-plane) or to $\xi_0 = \xi_- + (0.001,0,0)$, near the fixed point. For each $(\alpha,\sigma)$ on a $1000^2$ grid, the map \Eq{Q3DMap} is iterated $T = 5000$ times to eliminate transients. The orbit is declared to diverge (white region) if $|\xi_t| > \kappa_{max}$ for some $t \le T$, where $\kappa_{max}$ is the maximum of \Eq{kappa} over the studied parameter region ($\kappa_{max} = 3.237$ for \ref{SmallDeltaCase} and $4.632$ for \ref{LorenzLikeCase}). Otherwise the point $\xi_T$ is iterated up to $90$ more steps, checking for return time, defined as the first time, $p$, for which the distance
\beq{FirstReturn}
    \|\xi_{T+p} - \xi_T\| < 10^{-4}.
\eeq
Thus we find approximately periodic orbits up to period $90$; the period $p$ is indicated by the color map in \Fig{ResonantEx}. If an orbit is not periodic by this definition, the point is colored black or gray; we will discuss the dynamics in these regions in \Sec{NonresonantAttractors}.

\InsertFig{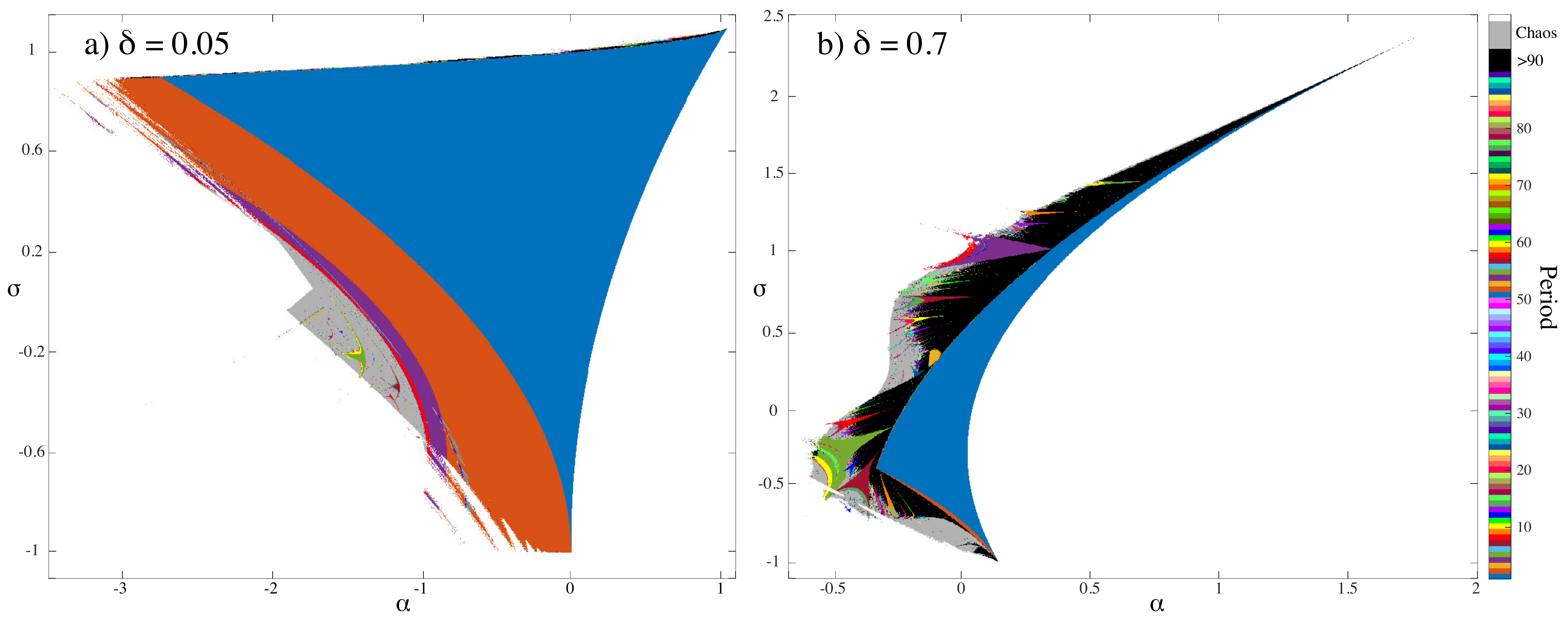}
{Parameter dependence of the dynamics of \Eq{Q3DMap} for (a) \ref{SmallDeltaCase} and (b) \ref{LorenzLikeCase} as $(\alpha, \sigma)$ vary. The white region corresponds to orbits that diverge. Bounded, periodic attractors, for periods up to $90$, are colored as shown in the color bar. 
The black region corresponds to period $>90$ or aperiodic, regular behavior, and the gray region to chaotic orbits, i.e., where the maximal Lyapunov exponent $\mu_T> \mu_o=0.0003$, see \Sec{NonresonantAttractors}.}{ResonantEx}{0.95}

The number of transient iterations, recurrence tolerance, and grid size were chosen to provide sufficient detail at the pictured resolution but still lessen the computational expense.

Note that since this method uses a single initial condition, it cannot find cases where there are multiple attractors. Additionally, it is possible that the chosen initial condition leads to an unbounded orbit when there is still an attractor somewhere in the $\kappa$-cube. Nevertheless, for almost all $(\alpha,\sigma)$ points that have bounded orbits in \Fig{BddVolumes}, the orbit iterated in \Fig{ResonantEx} is also bounded; therefore, with a few exceptions, it does not appear that the orbit of $\xi_0$ is unbounded when there are other bounded orbits. 
We plan to discuss such exceptional cases and the case of multiple attractors in future papers. 

The largest, blue regions in \Fig{ResonantEx} correspond to period-one, where $\xi_-$ is attracting. This region is bounded by the SN, PD, and NS curves seen in \Fig{FixPtBifs}. When the fixed point loses stability at the PD curve an attracting period-two orbit is born; the period-two (vivid orange) 
region is prominent in \Fig{ResonantEx}(a), though there is also a thin period-two region just below the PD curve in \Fig{ResonantEx}(b). Period-doubling bifurcations leading to period-four (magenta) and eight (red) are also seen in \Fig{ResonantEx}(a). By contrast, in \Fig{ResonantEx}(b), the period-two orbit in \ref{LorenzLikeCase} looses stability by
a NS bifurcation, so a doubling cascade is not seen in this case.

Resonant ``tongues'', analogous to the Arnold tongues found in circle maps, start along the NS curve where $\omega$ is rational; these are prominent in \Fig{ResonantEx}(b). Especially visible are the $\omega = \tfrac13$ (yellow), $\tfrac14$ (magenta), $\tfrac25$ (dark green), and $\tfrac37$ (brown) tongues. Bifurcation points along the NS curve for the first two of these were indicated in \Fig{FixPtBifs} as R3 and R4.


Enlargements of case~\ref{SmallDeltaCase}, shown in  \Fig{HenonCascadeTongue}, and of case~\ref{LorenzLikeCase}, \Fig{LorenzTongues}, provide more detail. In the following two subsections, we show a few examples of the corresponding orbits in phase space for the period-doubling cascade and resonant tongues.

\InsertFig{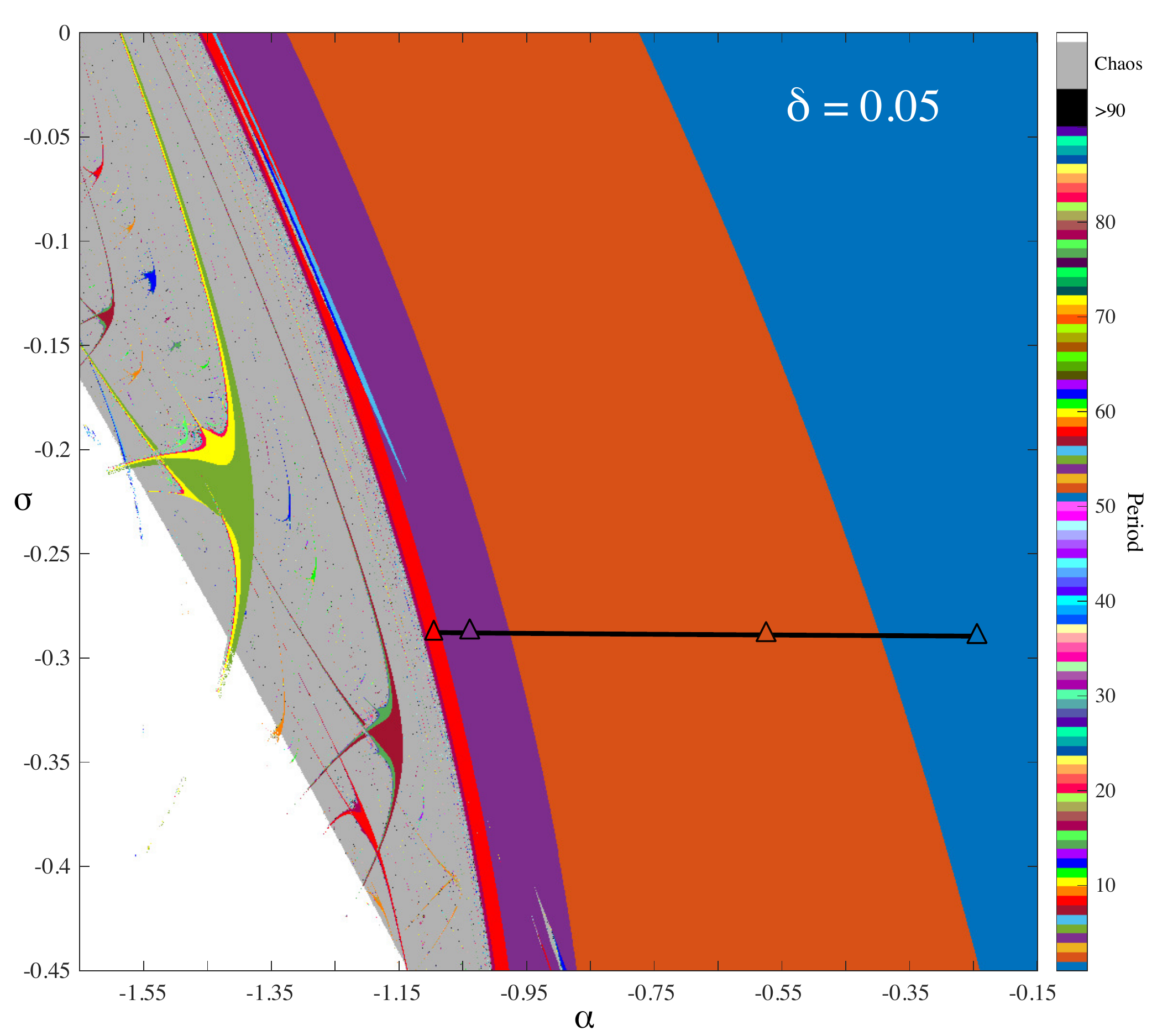}{An enlargement of case~\ref{SmallDeltaCase} from \Fig{ResonantEx}(a), providing more detail of the period-doubling cascade. The triangles along the line segment show parameters used in \Fig{HeteroclinicOrbits}. Also seen are a number of \textit{shrimps} corresponding to period-5 (dark green), period-7 (brown), etc. attracting orbits.}{HenonCascadeTongue}{0.9}

\subsection{Period-Doubling Cascade}\label{sec:Doubling}

A supercritical period doubling bifurcation generically occurs when a real multiplier of an attracting periodic orbit passes through $-1$, the PD line of \Fig{TSPlanes}, and results in the creation of a stable orbit of twice the period. 
As is well known from studies of 1D maps, this behavior often recurs, resulting in a cascade of doubling bifurcations that accumulates leading to the formation of a chaotic attractor.\cite{Guckenheimer02, Kuznetsov04}
Such a cascade is prominent in case~\ref{SmallDeltaCase} seen in \Fig{ResonantEx}(a) and its enlargement \Fig{HenonCascadeTongue}. Cascades are also often seen at the ends of the resonant tongues; most apparent for case~\ref{LorenzLikeCase} in \Fig{LorenzTongues}, which shows a multitude of tongues radiating from the attracting fixed point region.

To visualize the progression of the period-doubling cascade in case~\ref{SmallDeltaCase}, we fix $\sigma\approx -0.291$
along the line segment shown in  \Fig{HenonCascadeTongue} and vary $\alpha$. The four $(\alpha,\sigma)$ points (triangles) along this segment correspond to the parameters used to find the orbits seen in \Fig{HeteroclinicOrbits}, which are pictured using triangles of the same color.
Figure \ref{fig:HeteroclinicOrbits} also shows the heteroclinic orbit that lies on
the unstable manifold of the fixed point $\xi_-$ (black triangle), to the attracting orbit. The unstable manifold of the fixed point is traced out by $500$ iterates of a set of $100$ initial conditions starting in a ball of radius 0.01 about $\xi_-$.
The first point, at $\alpha\approx-0.248$, is a stable period-one orbit. It period doubles at $\alpha\approx-0.393$, creating an attracting period-2 orbit, shown for $\alpha \approx -0.563$ (orange triangles). The unstable manifold of $\xi_-$ is also shown in orange. As $\alpha$ continues to decrease the doubling recurs forming
next a stable period-4 (magenta) and then a stable period-8 (red).
The parameters for these are listed in \Tbl{OrbitParameters}. 

Also prominent in \Fig{HenonCascadeTongue} are a number of \textit{shrimps}, common structures seen in two-parameter families of one \cite{Gallas94,MacKay87f} and two-dimensional maps. \cite{Facanha13} Shrimps are prototypical structures
formed by saddle-node bifurcations of periodic orbits in a `sea of chaos.' These structures will not be discussed further in this paper.

\begin{table*}[ht]
\centering
\begin{tabular}{l|c|c|c}

    Case & Behavior & $(\alpha,\sigma)$ &  $\triangle$ Color \\
    \hline
    $\delta=0.05$ & fixed point & $(-0.24830,-0.29130)$ & black     \\
    & doubling bifurcation & $(-0.39312,-0.29130)$ &  \\
     & period-2 attractor & $(-0.56271,-0.29130)$ & orange  \\
     & quadrupling bifurcation & $(-0.97316,-0.29130)$ &  \\
     & period-4 attractor & $(-0.99417,-0.29130)$ & magenta  \\
     & octupling bifurcation & $(-1.08197,-0.29130)$ &  \\
     & period-8 attractor & $(-1.09922,-0.29130)$ & red  \\
    \hline
        $\delta=0.7$ & period-3 attractor & $(-0.13557,0.28892)$ & yellow    \\
     & period-4 attractor & $(0.10553,1.00950)$ & magenta  \\
\end{tabular}
\caption{Parameters for \Fig{HeteroclinicOrbits} and \Fig{LorenzHetero} that correspond with the annotations in \Fig{HenonCascadeTongue} and \Fig{LorenzTongues}.}
\label{tbl:OrbitParameters}
\end{table*}

\InsertFig{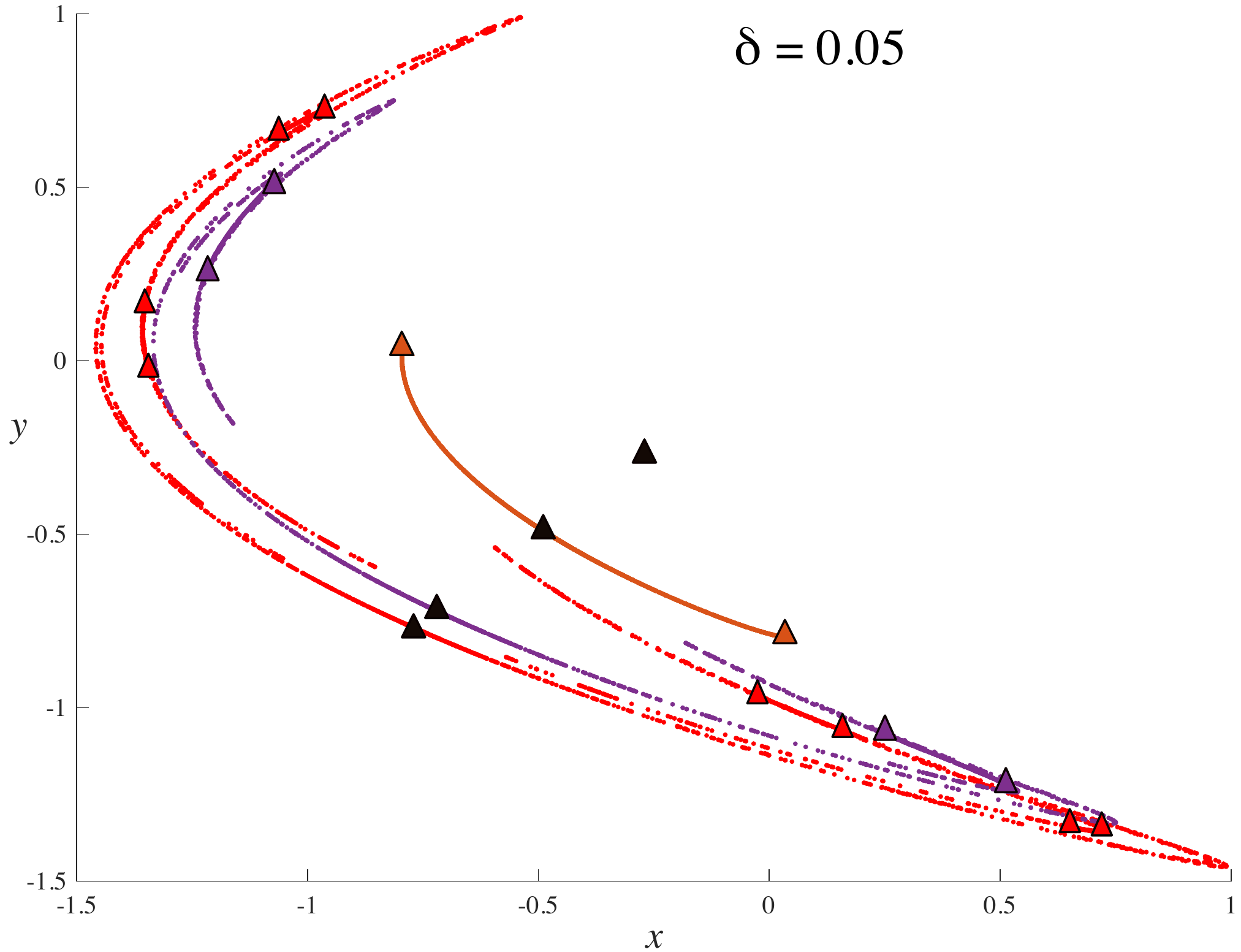}{
Four orbits along the period-doubling cascade for case~\ref{SmallDeltaCase}
for parameters given in \Tbl{OrbitParameters}. For each, the fixed point is shown as a black triangle, and attracting orbits as triangles colored by period, using the same color scheme of \Fig{HenonCascadeTongue}. At first, there is a single fixed point, and as $\alpha$ decreases, the fixed point shifts down and leftward, becoming a reflecting saddle when it bifurcates creating a period-2 attractor (orange). This process repeats for period-4 (magenta) and period-8 (red). A heteroclinic orbit of the same color connects the fixed point and attracting orbit.  
}
{HeteroclinicOrbits}{.5}

\subsection{Resonant Tongues}\label{sec:ResonantTongues}

\InsertFig{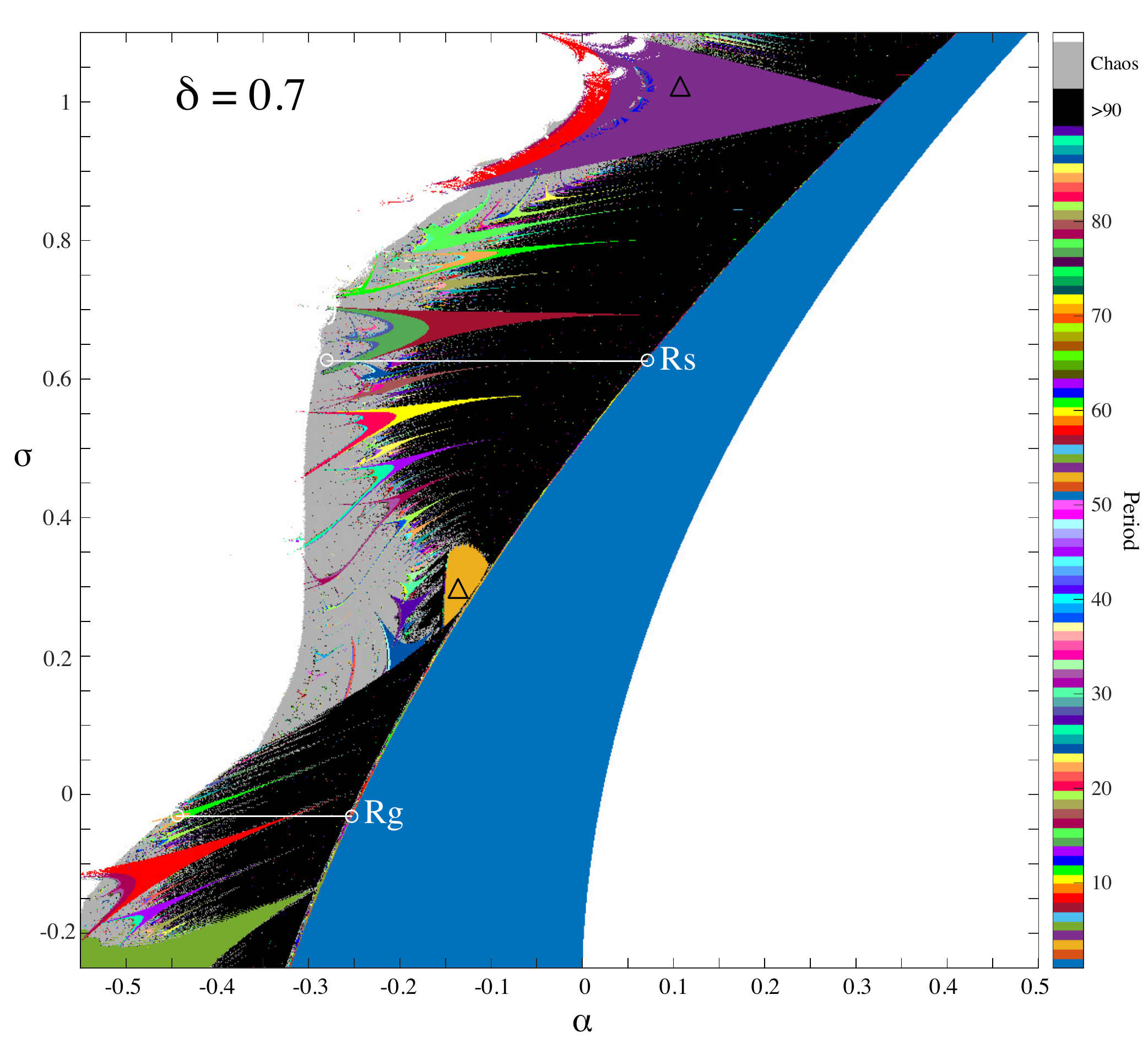}{An enlargement of case~\ref{LorenzLikeCase} from \Fig{ResonantEx}(b) provides detail of the resonant tongues emanating from the NS line of $\xi_-$. Highlighted points include two triangles in the period-3 and 4 tongues, used for the phase portraits in \Fig{LorenzHetero}. Also shown are line segments starting at $(\alpha_g,\sigma_g)$ and $(\alpha_s,\sigma_s)$ along the NS line. These correspond to $\omega=2/(1+\sqrt{5})$, the golden mean, and $\omega=1/\sqrt{2}$, the silver mean. The behavior of orbits in phase space along these segments are discussed in \Sec{Nonchaotic}.}{LorenzTongues}{0.9}

At a resonant point, $\omega = \tfrac{p}{q}$, on the NS curve, a pair of period $q$ orbits are born when $q \ge 5$. These period-$q$ orbits exist
in a ``tongue''-shaped region bounded by curves of SN bifurcations. While a pair of periodic orbits still exist for the strongly resonant cases, $q \le 4$, the resonant regions can be more complex\cite{Kuznetsov04}.
Near the NS curve of the fixed point, one of these orbits is stable and is easily detected by our recurrence algorithm. Such tongues are especially prominent in \Fig{LorenzTongues} for case~\ref{LorenzLikeCase}. 

The pair of orbits found within the tongues can be used to find heteroclinic orbits from the period-$q$ saddle to its attracting counterpart. Examples for $q= 3$ and 4 are seen for case~\ref{LorenzLikeCase} in \Fig{LorenzHetero}. Parameters are chosen in the respective tongues are indicated in \Fig{LorenzTongues} by triangles and listed in \Tbl{OrbitParameters}. To visualize the heteroclinic orbits, a set of $100$ initial conditions are chosen in a ball of radius 0.01 about a point on the period-$q$ saddle, which is found by numerically solving for a fixed point of $L^q(\xi)$. The first $500$ iterates of these points, shown in \Fig{LorenzHetero}, trace out the unstable manifold of the saddle, and lie on the stable manifold of the period-$q$ attractor.

\InsertFigTwo{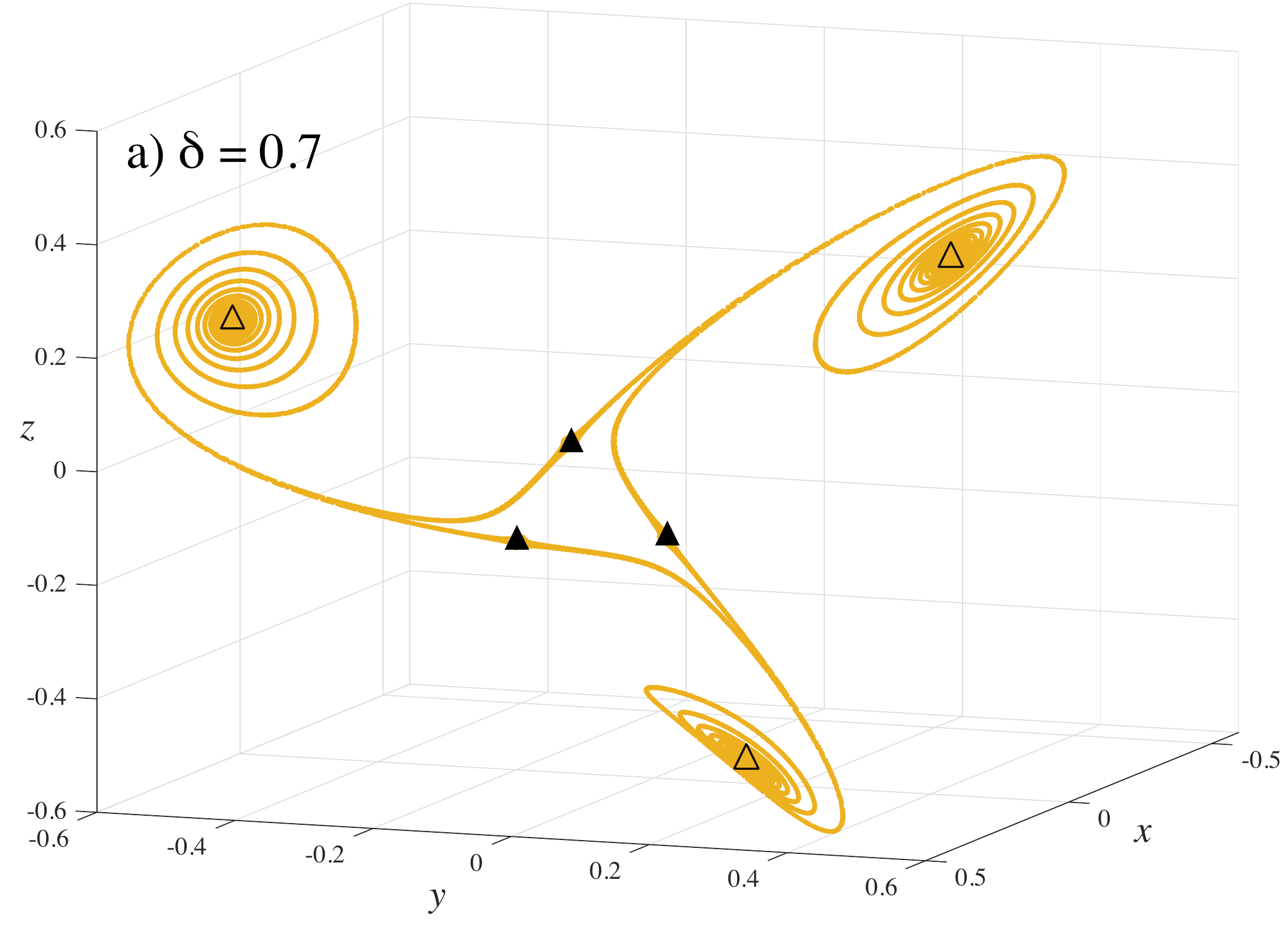}{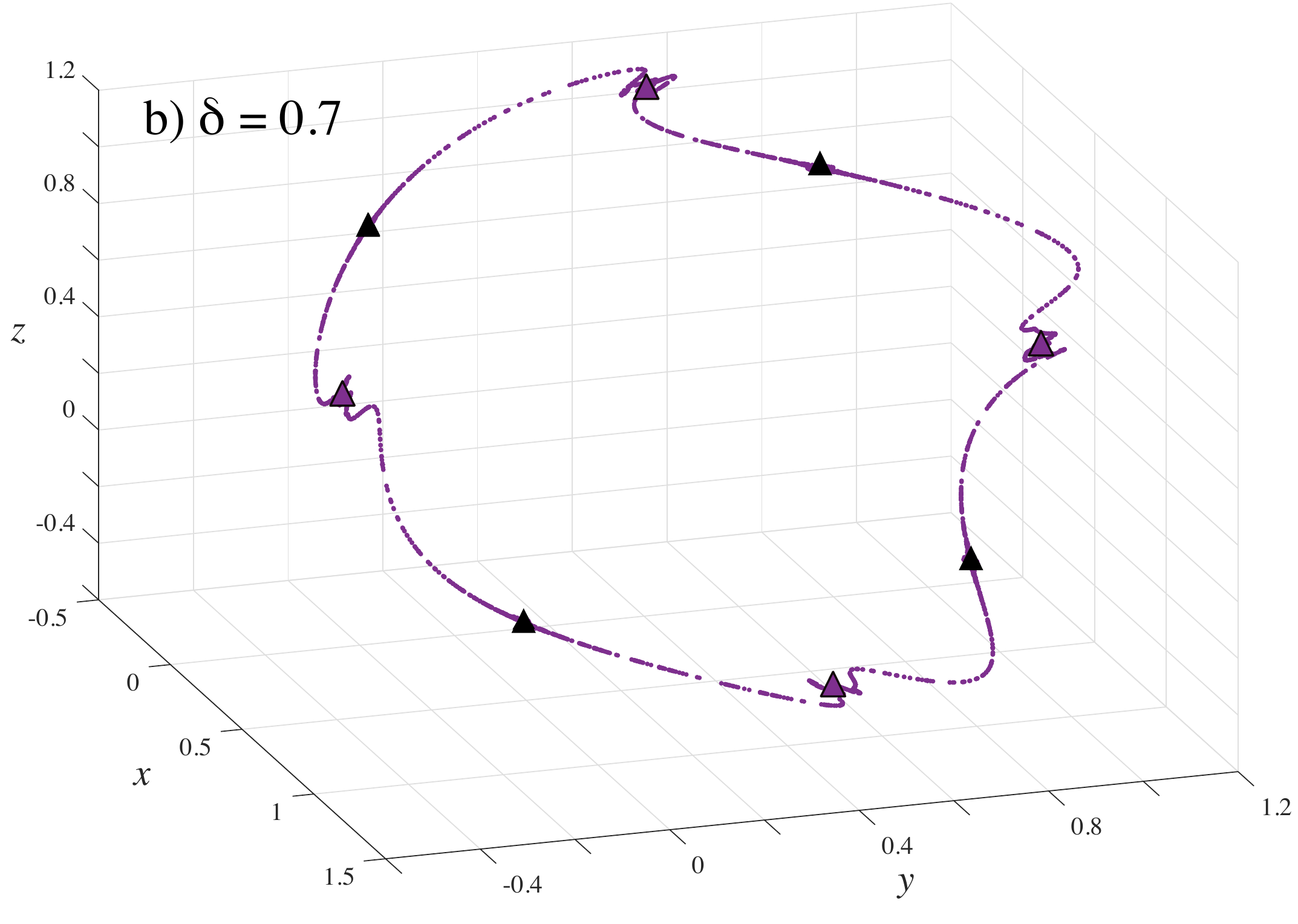}{For the \ref{LorenzLikeCase} case, heteroclinic orbits are found in the period-3 (yellow, left) and 4 (magenta, right) tongues respectively. See \Tbl{OrbitParameters} for values of $(\alpha,\sigma)$. 
}
{LorenzHetero}{.5}

\section{Aperiodic Attractors} \label{sec:NonresonantAttractors}

We now discuss the black and gray regions seen in \Fig{ResonantEx}, and its enlargements \Fig{HenonCascadeTongue} and \Fig{LorenzTongues}. These represent parameters for which there is an attractor that is either aperiodic or periodic with period larger than $90$. Possible aperiodic attractors include invariant circles formed at an NS bifurcation with irrational $\omega$, as well as more complex, possibly chaotic cases. In this section, we will distinguish between regular and  chaotic cases by computing the maximal Lyapunov exponent. In a number of studies, multiple Lyapunov exponents were calculated to characterize the dimensionality of the unstable spaces of attractors.\cite{Gonchenko21a,Gonchenko16} Here we only compute one, as our focus is simply the distinction between chaotic and regular.

Formally, the Lyapunov exponent is
\beq{MLE}
    \mu(\xi_0;v_0) =\limsup_{n\to\infty} \frac{\ln \|v_n \|}{n} .
\eeq
for an initial point $\xi_0$, and an initial vector $v_0$, which evolves linearly as $v_t=DL(\xi_{t-1})v_{t-1}$. Note for a generic initial vector, $v_0$, this limits to the maximal exponent (MLE). If there exists an attracting regular orbit, then any initial condition in its basin will have a non-positive MLE, whereas chaotic orbits will have $\mu > 0$.

Numerically, we approximate the $\limsup$ of \Eq{MLE} by choosing an increment $\Delta T$, and computing
\beq{MLENumeric}
    \mu_T(\xi_0;v_0) = \max_{n\in [T,T+\Delta T)} \frac{\ln \|v_n \|}{n} .
\eeq
This approximates \Eq{MLE} for a ``large'' enough $T$ and $\Delta T$, up to some tolerance. 
We estimate the MLEs for those bounded orbits that have periods larger than 90. The
initial point used for \Eq{MLENumeric} is the point used in the recurrence algorithm described in \Sec{PeriodicAttractors} found after $5000+90$ iterates, and the initial vector is set to $v_0=(1,1,1)/\sqrt{3}$.  To avoid overflow, the length of $v_n$ is renormalized every 10 iterates. As is well known, it is computationally expensive to achieve convergence of MLEs. Indeed, since the ostensible error for \Eq{MLENumeric} is $\cO(T^{-1})$, we chose a threshold appropriate for $T \sim 10^3-10^4$: an orbit is deemed regular if, $\mu_T \le \mu_o = 3(10)^{-4}$.
We selected the interval $\Delta T = 100$ as it gave reasonable convergence for a number of trials. 
The time $T$ in \Eq{MLENumeric} is increased in steps of $\Delta T$ until the error $|\mu_T -\mu_{T+\Delta T}| < 10^{-4}$, or until $T$ reaches $T_{max} = 10^5$. In this case, the MLE is set to  $\mu_{T_{max}}$. For the results in \Fig{ResonantEx}, $\mu_T$ is set to $\mu_{T_{max}}$ for $0.14\%$ of the calculations for case~\ref{LorenzLikeCase} and $1.2\%$ for case~\ref{SmallDeltaCase}. 

Orbits with $\mu_T > \mu_o$ are declared to be ``chaotic'', and colored gray, and those with smaller MLEs are declared ``regular'' and colored black in the figures. As seen in \Fig{ResonantEx}. there is a qualitative difference between the two cases.  For case~\ref{SmallDeltaCase}, there are very few regular, aperiodic orbits--- these are confined to a narrow strip along the NS curve at the top of the bounded region ($4.3\%$ of the aperiodic orbits). For case~\ref{LorenzLikeCase}, however, there are large regions of regular, aperiodic behavior to the left of the NS curve ($70\%$ of the aperiodic orbits). In the following subsections, we illustrate this difference by looking at the corresponding behavior of orbits in phase space. 

\subsection{Regular Attractors}\label{sec:Nonchaotic}

Regular, aperiodic orbits are primarily born on the NS curve when the rotation number, $\omega$, in \Eq{NSEvals}, is irrational.  Such a bifurcation generically gives rise to an invariant circle that persists for an interval along a curve in parameter space that starts on the NS curve. \cite{Kuznetsov04} The rotation number will become rational when the curve enters a resonant tongue. To explore the structure of such circles, we study the development of orbits in phase space for \ref{LorenzLikeCase} along two lines in the $(\alpha,\sigma)$-plane that start at an irrational NS point, as illustrated in \Fig{LorenzTongues}:
\begin{widetext}  
\begin{align}
    &\text{Rg}:  &\sigma_g &\approx -0.03232,  
    &(\alpha_{g0},\alpha_{g1}) &\approx (-0.25375,-0.45269), \label{eq:GoldenSegment}\\
    &\text{Rs}:  &\sigma_s  &\approx 0.62724,   
    &(\alpha_{s0},\alpha_{s1}) &\approx (0.07064, -0.28960)
    \label{eq:SilverSegment}.
\end{align}
\end{widetext}
The Rg segment starts at $(\alpha_{g0},\sigma_g$), the NS bifurcation for the golden mean, $\omega=\tfrac12 (\sqrt{5}-1)\approx0.61803$, and the Rs segment starts at $(\alpha_{s0},\sigma_s)$, the NS bifurcation for the silver mean, $\omega = \tfrac{1}{\sqrt{2}}\approx0.70711$. Each segment has fixed $\sigma$ and ends when $\alpha$ reaches the edge of the bounded region.

Figure~\ref{fig:LorenzGoldenOrbitDevelopment} shows orbits along \Eq{GoldenSegment}, starting just below $\alpha_{g0}$ and moving towards $\alpha_{g1}$ in six panels. These are projected onto the plane orthogonal to the line $x=y=z$ that contains the fixed points. This plane is spanned by the orthogonal vectors
\beq{UVPlane}
    u=(-1,1,0), \quad v=(1,1,-2),
\eeq 
and these are used as the axes in the figures.
We observe that the invariant circles appear to have a one-to-one projection onto this plane and enclose the projected fixed point $\xi_-$, which projects to the origin (black triangle). Each panel consists of several orbits; corresponding values of $\alpha$ are given. As $\alpha$ decreases, 
an invariant circle grows, deforms, and bifurcates to periodic orbits when the segment passes through resonant tongues.

An alternative visualization of the dynamics along \Eq{GoldenSegment} is through its bifurcation diagram, seen in \Fig{IrrationalInfo}(top, left). This shows a 1D projection
onto the $x$-axis for varying $\alpha$.
Invariant circles correspond to dense segments and periodic attractors to isolated points in this diagram. 

\InsertFigSixH{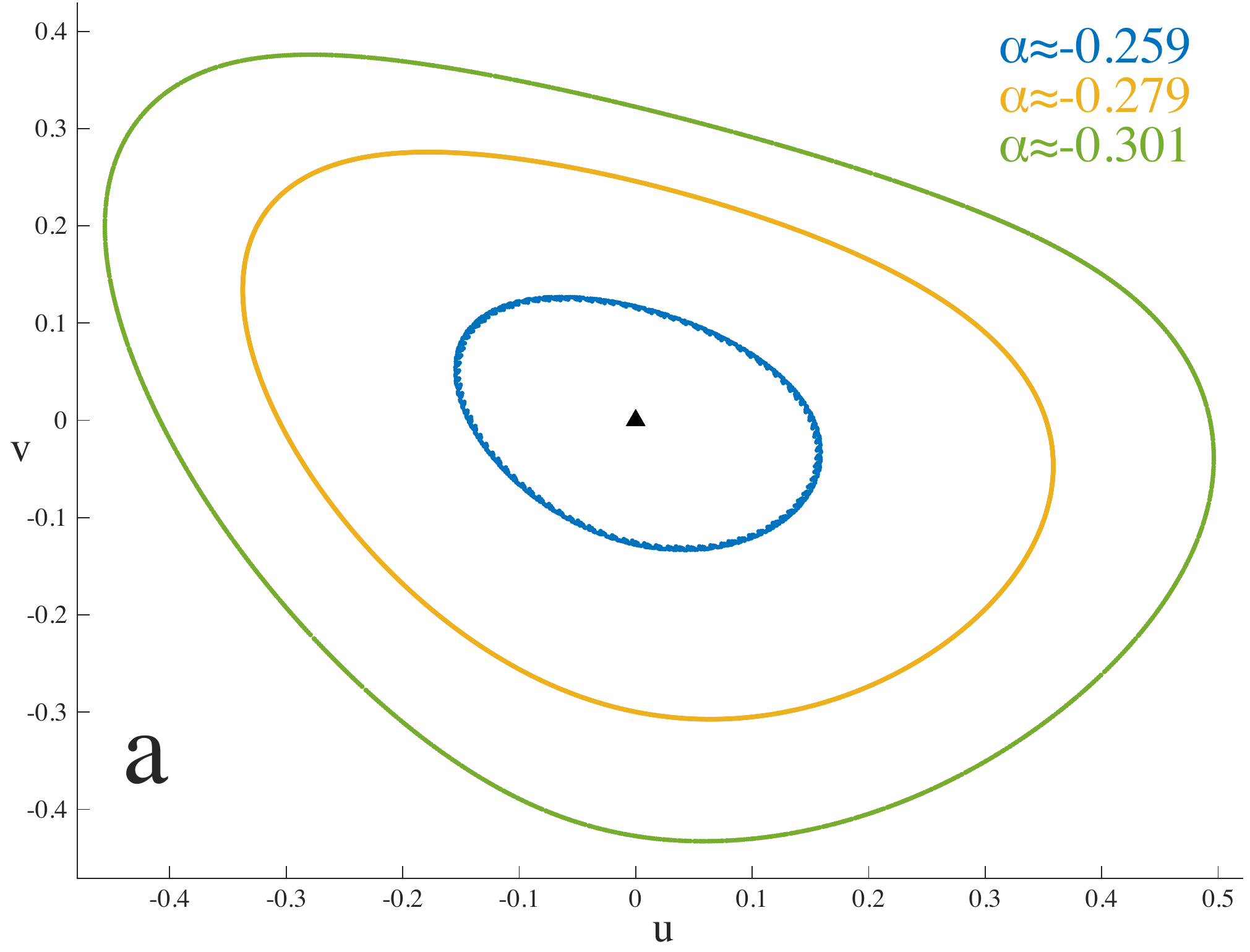}{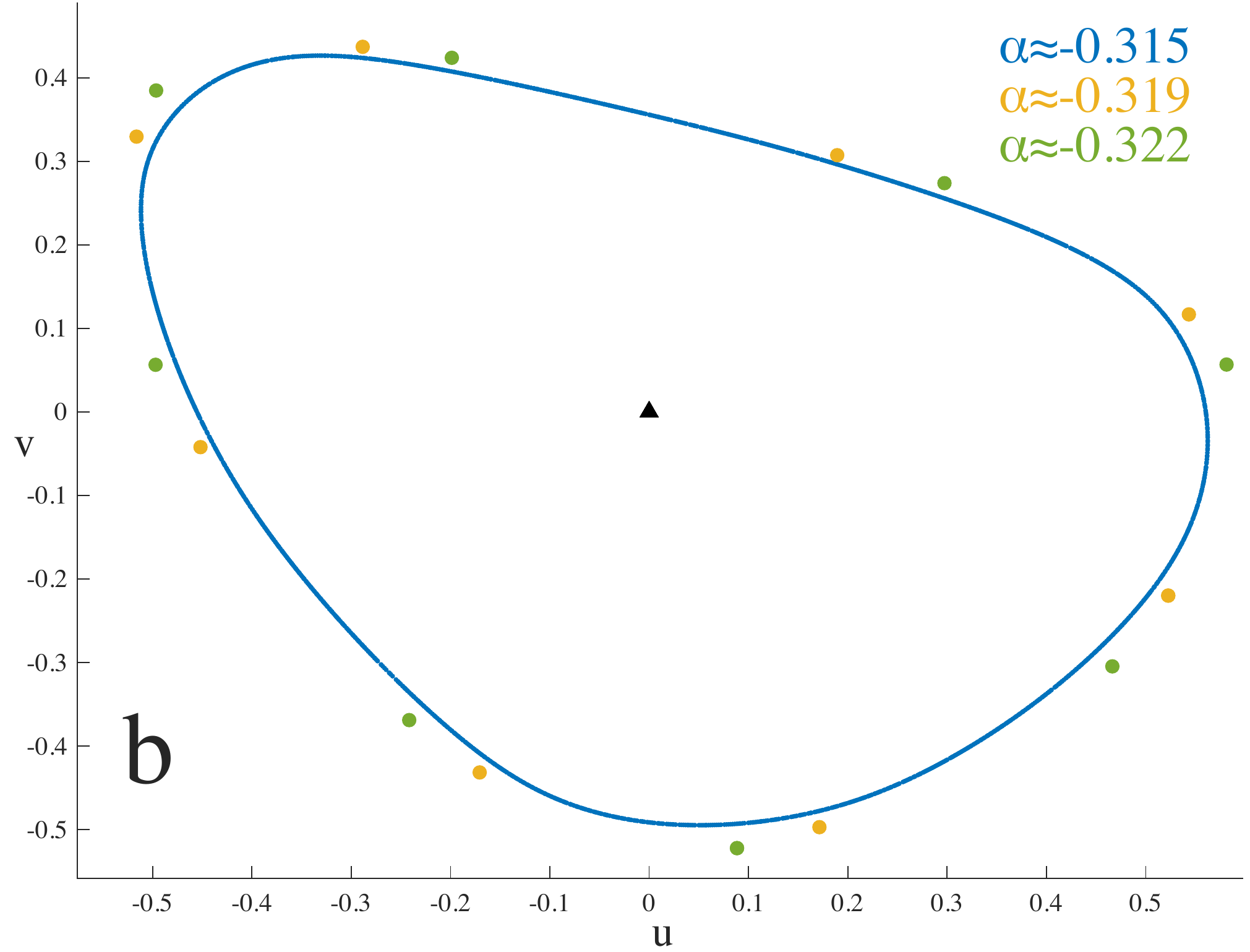}{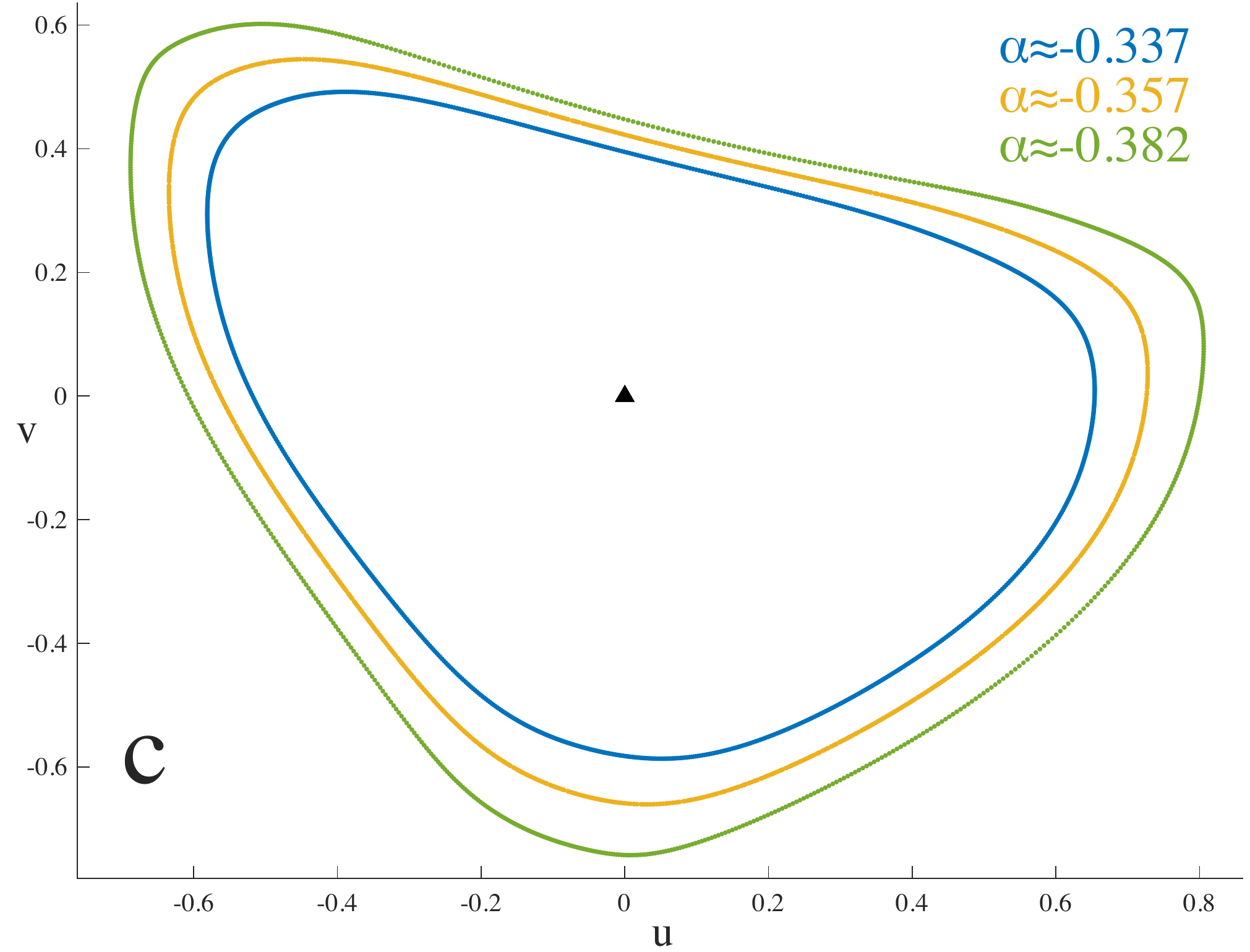}{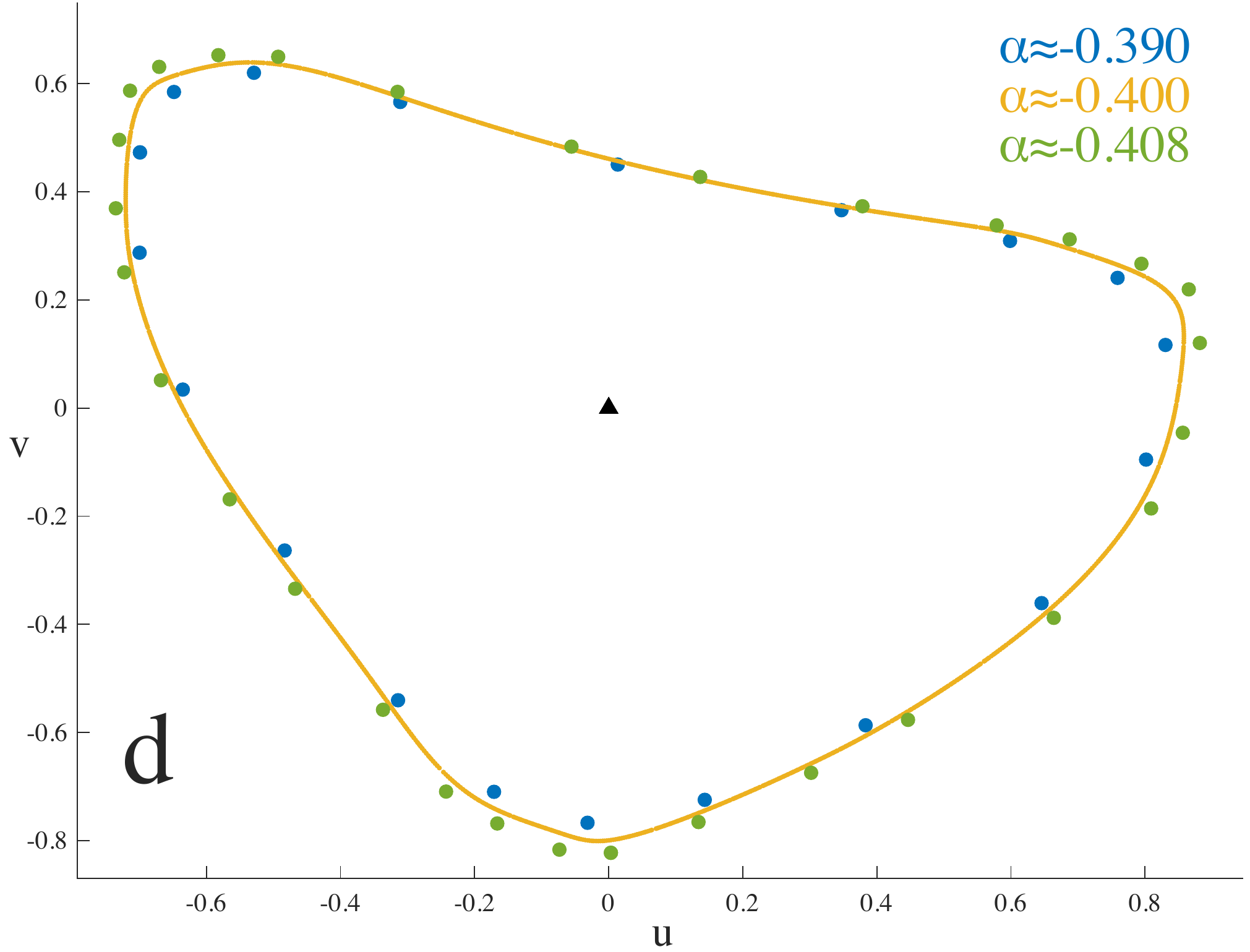}{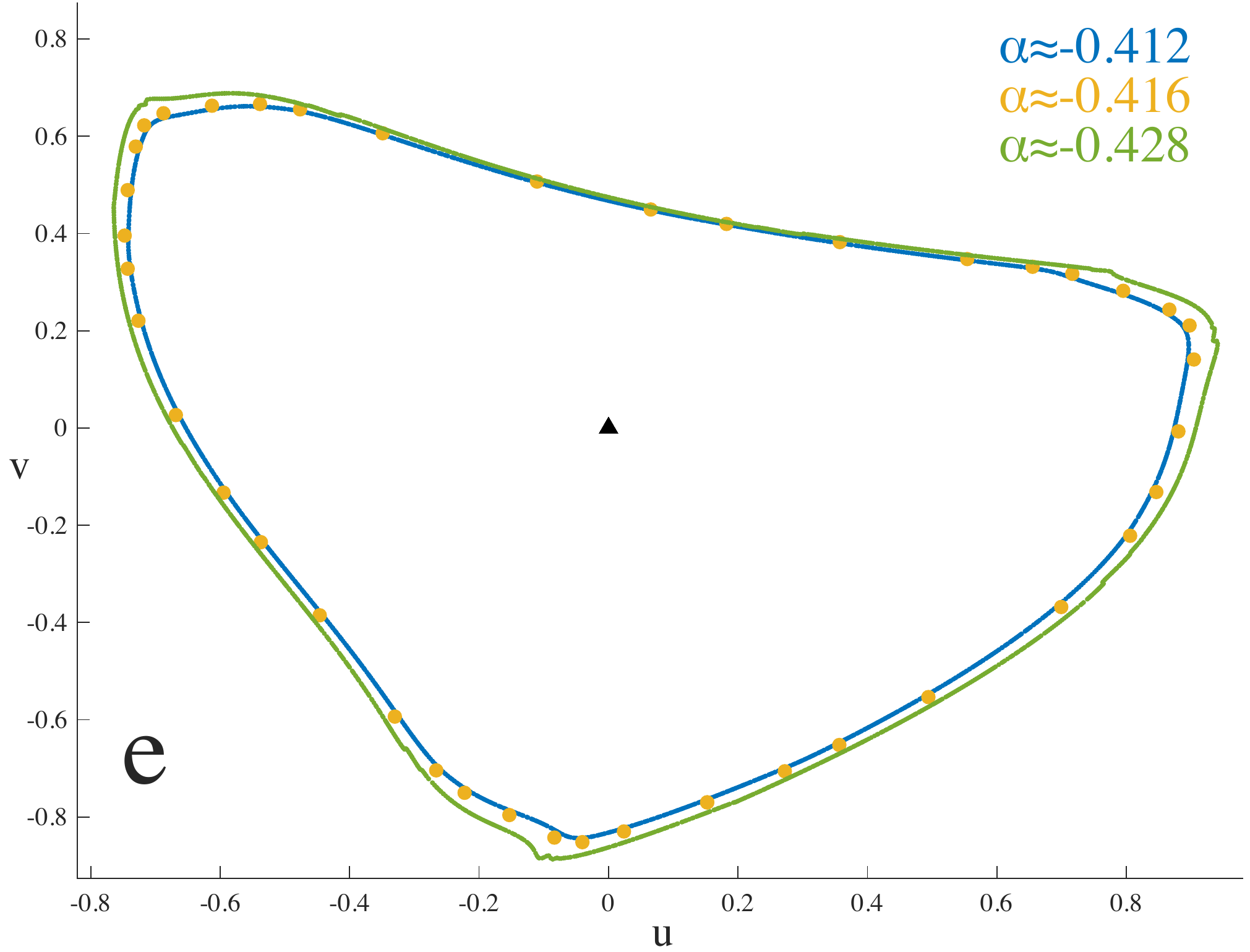}{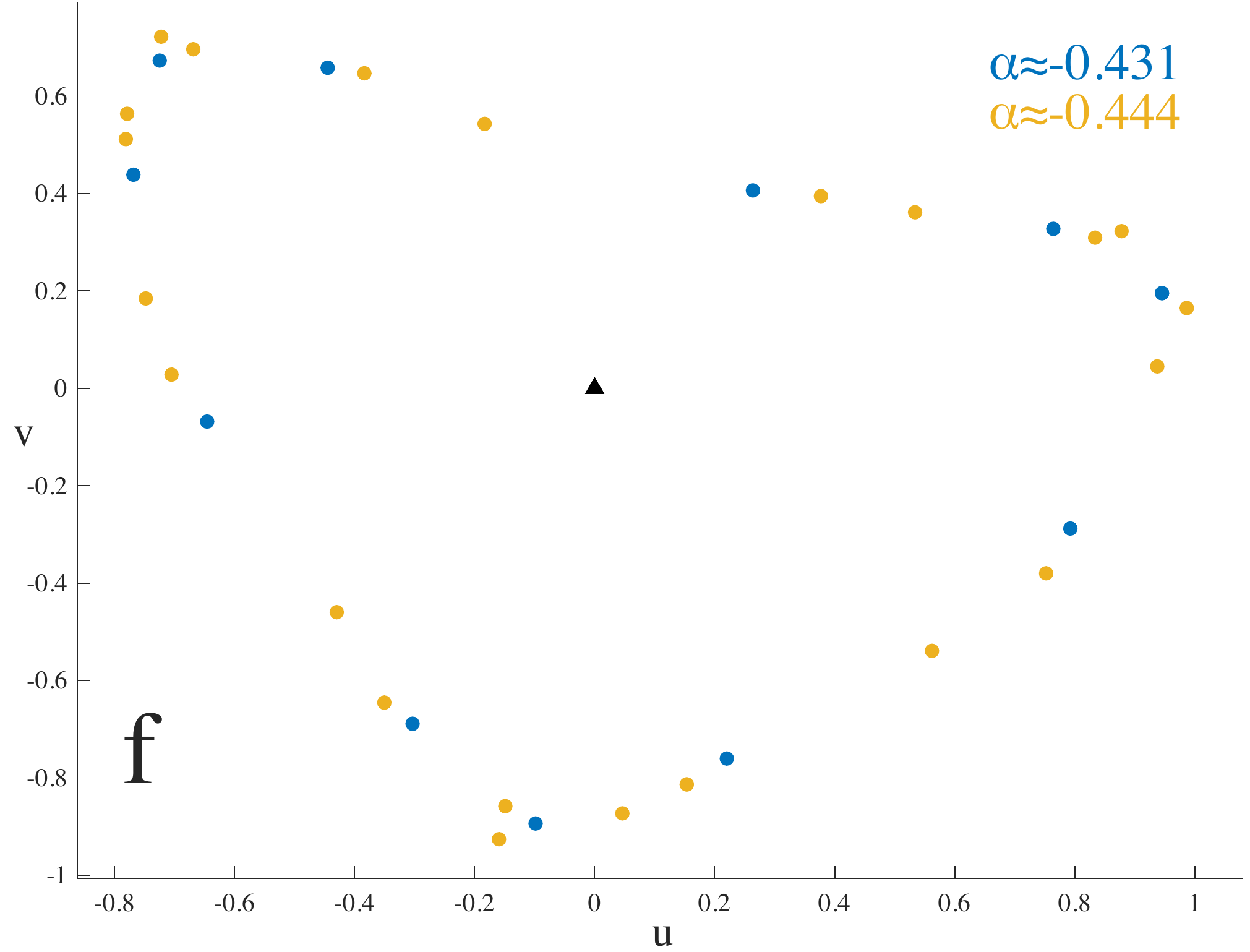}{Attracting orbits found along \Eq{GoldenSegment}, seen in \Fig{LorenzTongues}, projected onto the plane \Eq{UVPlane}. The black triangle at the origin is the projected fixed point, $\xi_-$. Each panel contains two or three orbits for values of $\alpha$ shown in the same color as the orbit. The panels correspond to the  intervals at the bottom of \Fig{IrrationalInfo}(left).
(a) Invariant circles close to $\alpha_{g0}$. 
(b) An invariant circle and a period-8 orbit. 
(c) Three invariant circles. 
(d) Period-19 and period-30 orbits, with an invariant circle at an intermediate $\alpha$. These periodic orbits lie in windows too small to be visible in \Fig{IrrationalInfo}(left). 
(e) Two invariant circles and a period-41 orbit at an intermediate $\alpha$. 
(f) Period-11 and doubled period-22 orbits. 
All of these orbits are regular, as $\mu_T<\mu_o$ along the entire segment, seen in \Fig{IrrationalInfo}(left). Orbits become divergent when $\alpha \leq \alpha_{g1}$.}{LorenzGoldenOrbitDevelopment}{0.3}

Following the analysis of \citeInline{Dullin09a}, we compute the rotation number of the orbits. Since the projection onto the $(u,v)$ plane is one-to-one and encircles the origin, the rotation number can be computed by measuring the angle $\theta_t$ at time $t$ counterclockwise from the vector $u$ using the full range atan2 function. The time $T$ approximation of the rotation number is then
\beq{RotationNumber}
    \omega_T=\frac{1}{2\pi T}\sum^T_{t=1}\theta_t.
\eeq
This sum is computed from the initial conditions used for \Fig{LorenzTongues},
after removing the transient as before. We choose $T$ to be the return time \Eq{FirstReturn}, or, if there is no such time, $T = 5(10)^5$. 

\InsertFig{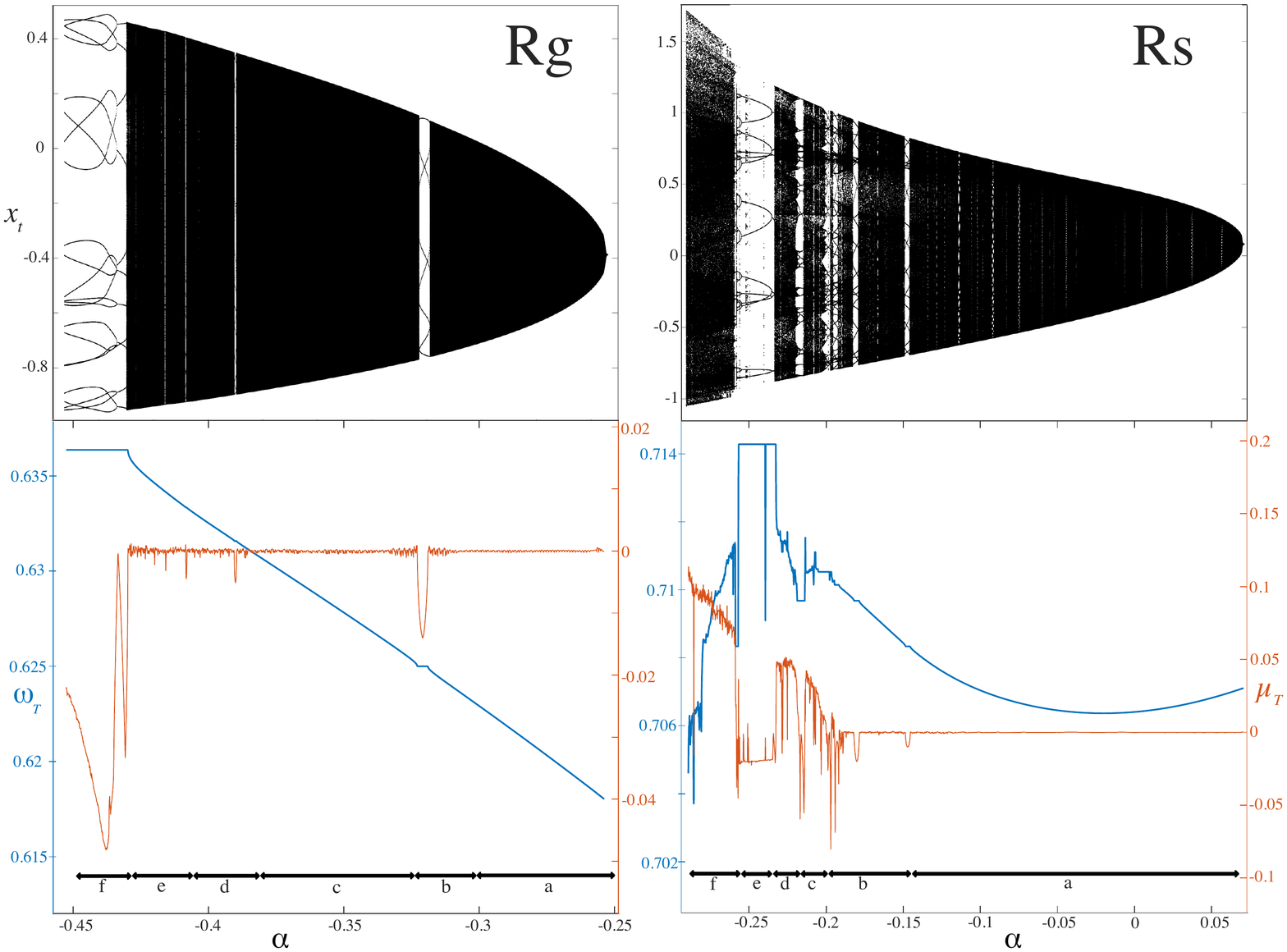}{Bifurcation diagrams (top) aligned with graphs of $\omega_T$ and $\mu_T$ (bottom) for attracting orbits along the segments \Eq{GoldenSegment}(left) and \Eq{SilverSegment}(right). Calculations start with an $\alpha$ value $10^{-4}$ below the NS points, and $\alpha$ is decreased until the orbits diverge (i.e., when the segment reaches the white region). Resonant tongues correspond to constant $\omega_T$ in the bottom panels and to windows of periodicity in the top panels. Chaotic orbits, $\mu_T > \mu_o$ are only seen for \Eq{SilverSegment}. The annotated $\alpha$ intervals, a-f, refer to the panels of \Fig{LorenzGoldenOrbitDevelopment} and \Fig{LorenzSilverOrbitDevelopment}.}{IrrationalInfo}{0.9}

The rotation number and corresponding Lyapunov exponent, \Eq{MLENumeric}, are shown as a function of $\alpha$ in the bottom row of \Fig{IrrationalInfo}, aligned with the corresponding bifurcation diagrams for \Eq{GoldenSegment} and \Eq{SilverSegment}.

We do not observe chaotic orbits along \Eq{GoldenSegment}---$\mu_T$ is essentially nonpositive, though small oscillations up to the threshold $\mu_o$ reflect the difficulty in computing the MLE. Note that when the dynamics are conjugate to a rigid rotation (i.e., the orbit lies on a circle), then there is a zero Lyapunov exponent. Since the circle is attracting, this is what we observe for the MLE in \Fig{LorenzTongues}.  When  $\alpha\approx\alpha_{g0}$, the rotation number $\omega_T\approx0.618$, as expected, and as $\alpha$ decreases, the rotation number grows monotonically. When the orbit passes through a resonant tongue, $\omega_T$ becomes a constant rational value and $\mu_T < \mu_o$ since the resulting periodic orbit is attracting. These align with the periodic windows in the bifurcation diagram. Periodic windows visible in \Fig{IrrationalInfo}(left) are the period eight ($\omega=\tfrac{5}{8}=0.625$) and the period 11 and 22 tongues. For the latter, $\omega=\tfrac{7}{11} \approx 0.636$, and---when the orbit doubles---$\omega = \tfrac{14}{22}$, the same value. Near the period doubling in the $\tfrac{7}{11}$ tongue,  $\mu_T$ grows as $\alpha$ decreases, reaching zero at the bifurcation point $\alpha \approx -0.433$.
The orbit diverges when $\alpha \le \alpha_{g1}$.  

The segment \Eq{SilverSegment} crosses multiple tongues and chaotic regions, as seen in \Fig{IrrationalInfo}(right). When $\alpha\approx\alpha_s$, $\omega_T\approx0.707$, but as $\alpha$ decreases the rotation number is not monotone. Note that when $\mu_T>\mu_o$, calculations for $\omega_T$ do not converge well since the orbit is chaotic. As before, in the resonant tongues, $\mu_T<\mu_o$  and $\omega_T$ is constant, and when there is an attracting invariant circle, $\mu_T$ is close to zero. The most visible tongues correspond to orbits of period 24 ($\omega=\tfrac{17}{24} \approx0.708$), 31 and 62 ($\omega=\tfrac{22}{31}=\tfrac{44}{62} \approx0.710$), and 7 and 14 ($\omega=\tfrac{5}{7}=\tfrac{10}{14} \approx0.714$). Note that there are chaotic regions both before and after the last tongue. 

Phase portraits along \Eq{SilverSegment} projected onto the plane \Eq{UVPlane}
are shown in a series of six panels in \Fig{LorenzSilverOrbitDevelopment}; 
each corresponds to an interval annotated in \Fig{IrrationalInfo}(right). 
Near $\alpha_{s0}$, we observe a family of growing invariant circles that pass through windows of periodicity as $\alpha$ decreases. Panels (c),(d), and (f) show chaotic attractors; these occur when the segment \Eq{SilverSegment} passes through gray regions, so that $\mu_T > \mu_o$, in \Fig{IrrationalInfo}(right). These attractors are born with a structure like the invariant circle, but also seem to fold around the neighboring, no-longer stable periodic orbits.
In panel (f) the outer boundary of the chaotic attractor appears relatively smooth, and aligns with the former invariant circle, but its interior is much more complex. This attractor is also shown in 3D in \Fig{LLChaoticAttractors}(a).

\InsertFigSixH{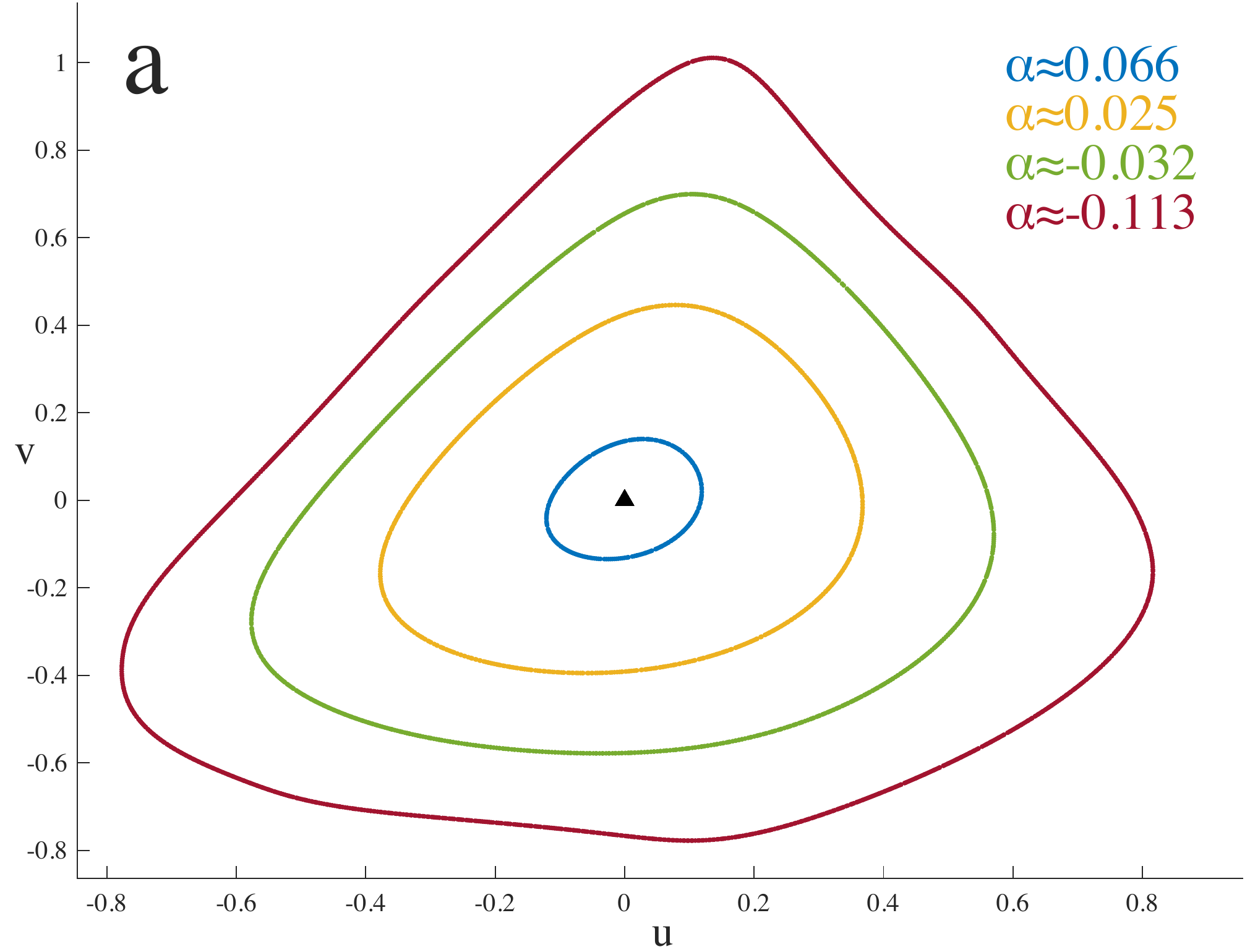}{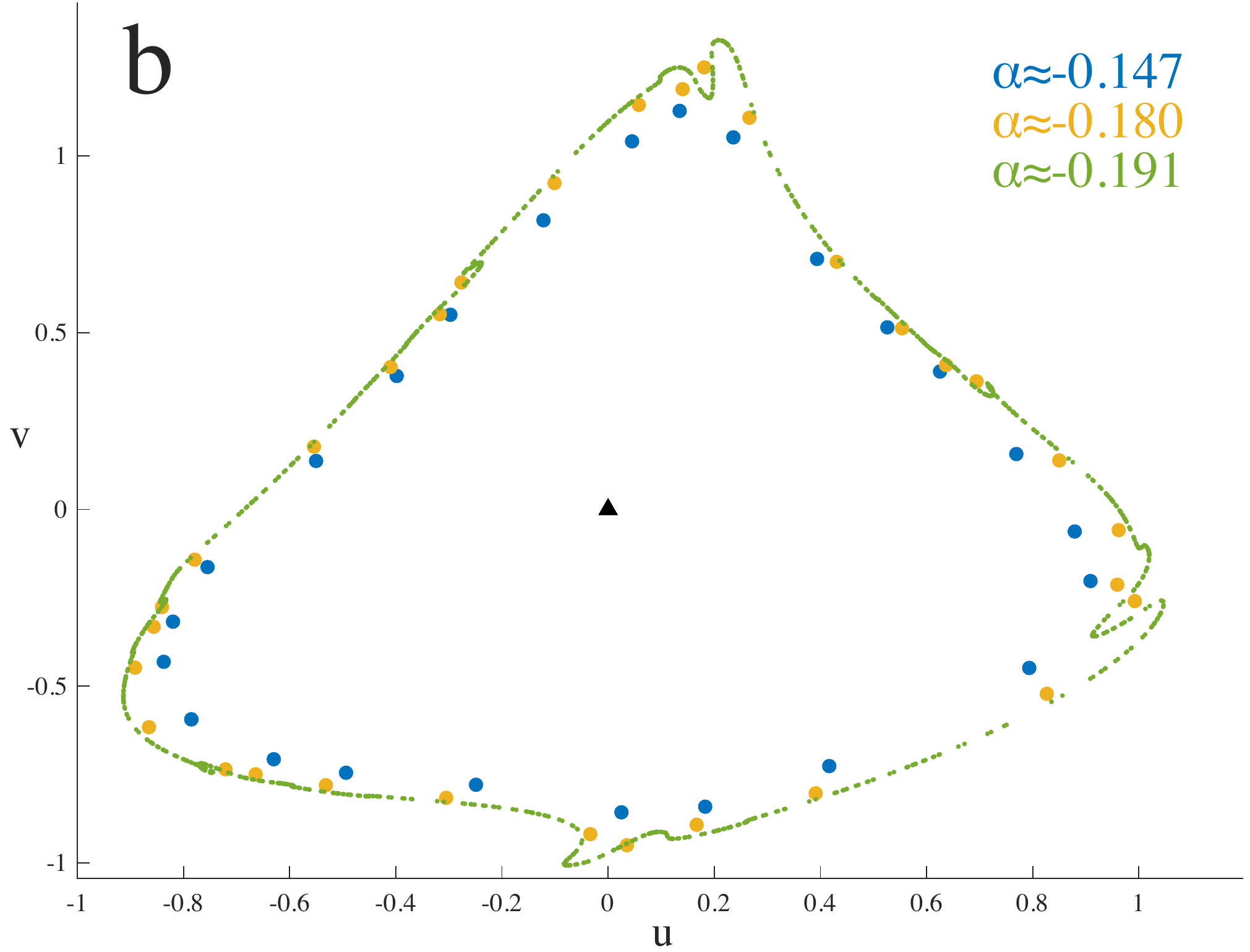}{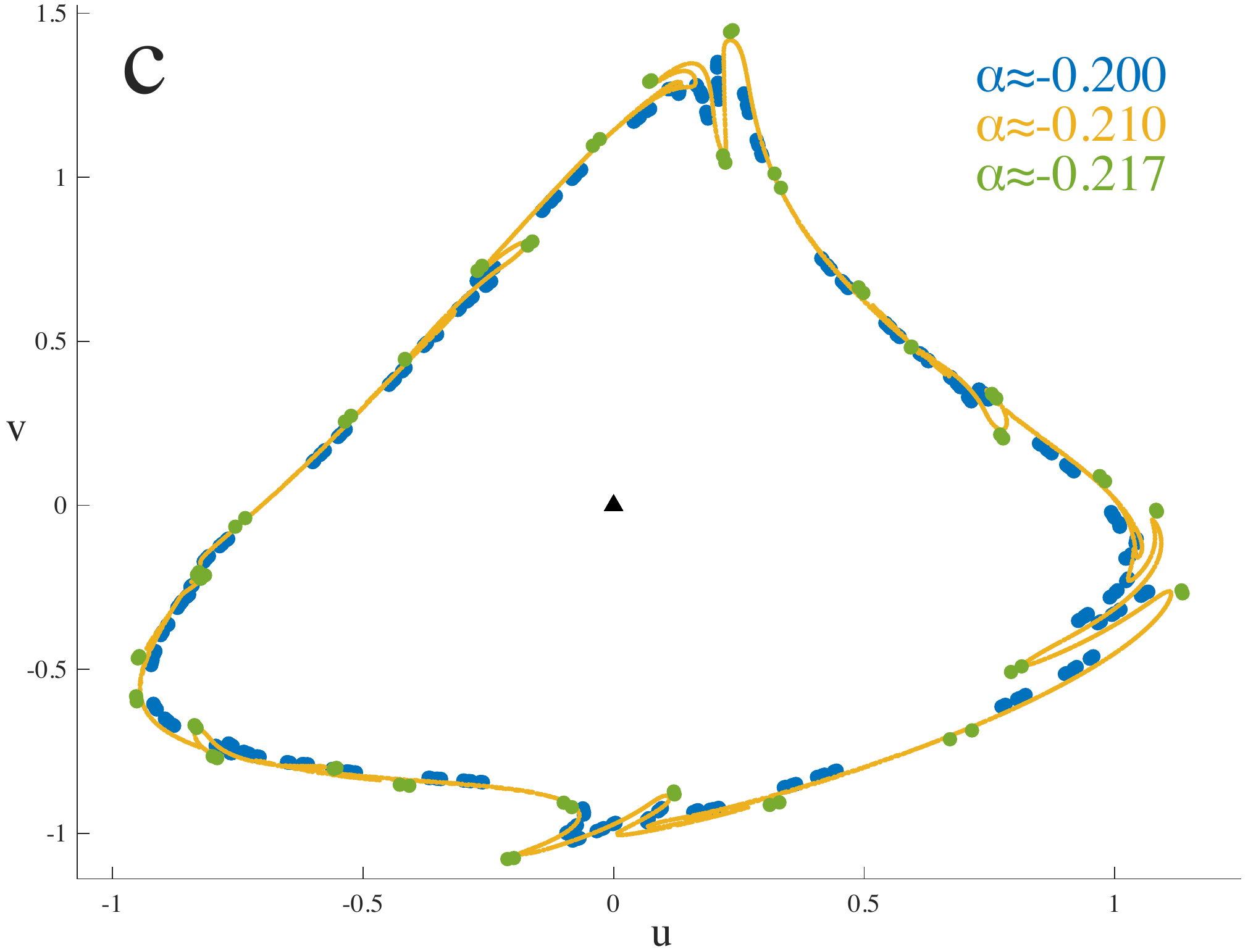}{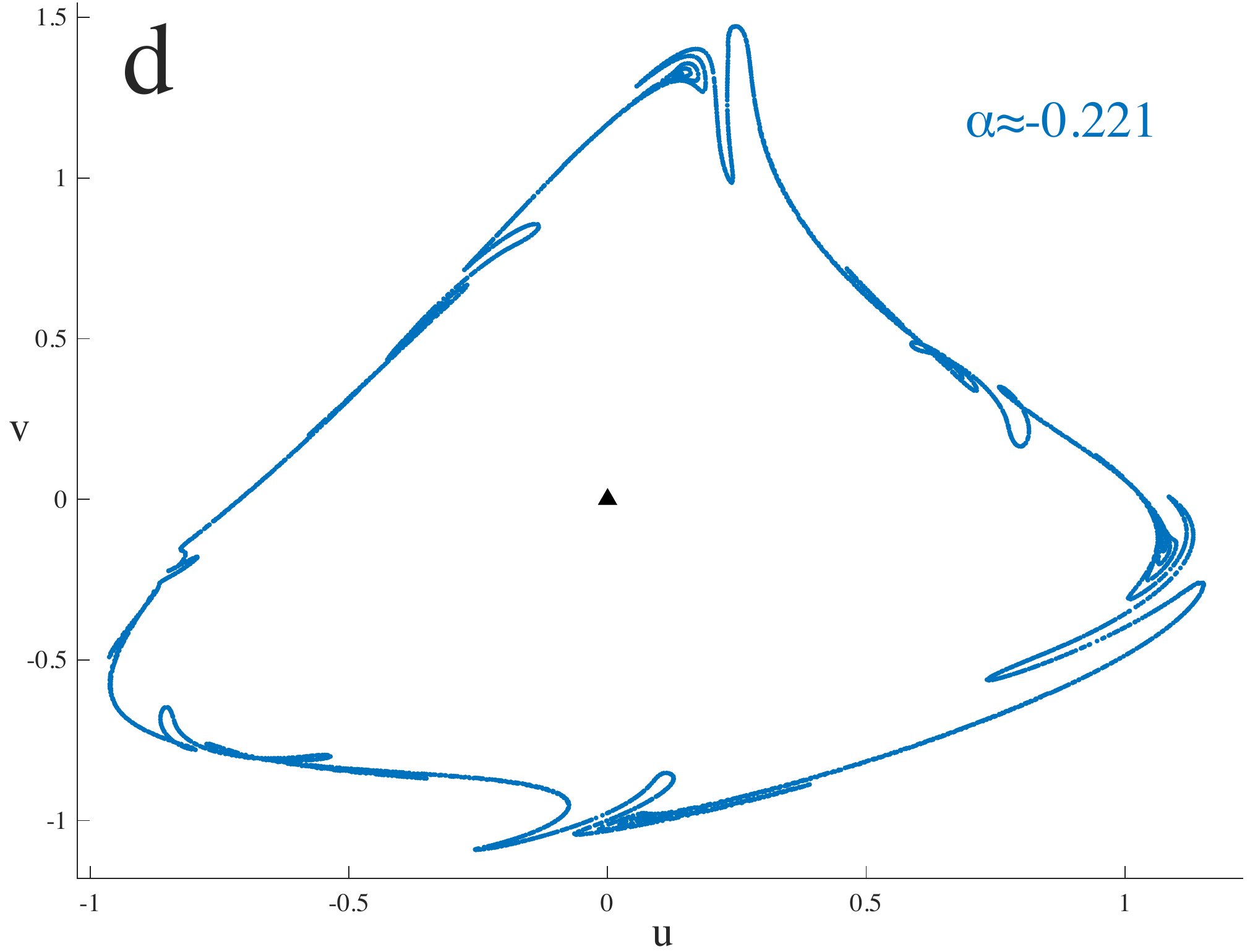}{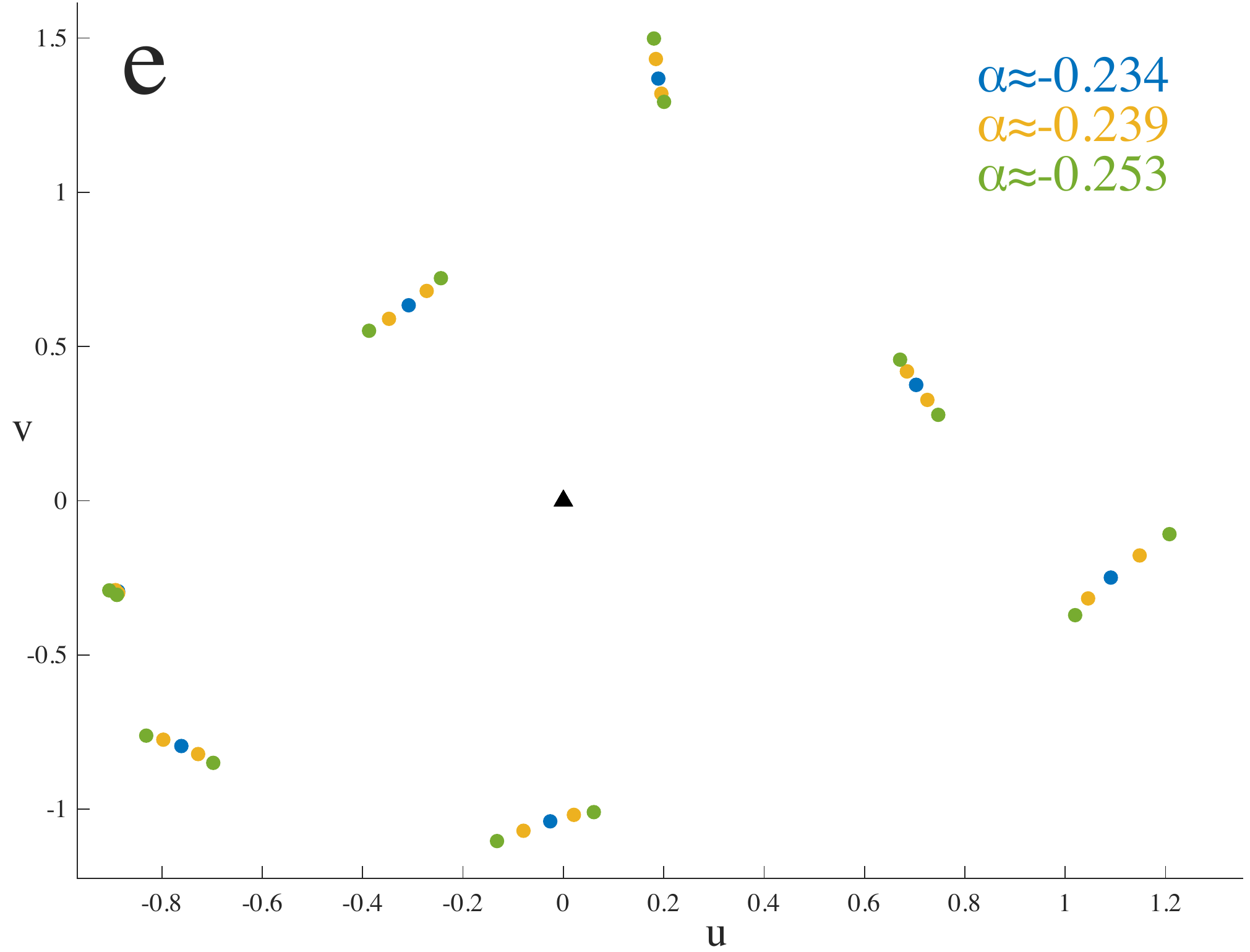}{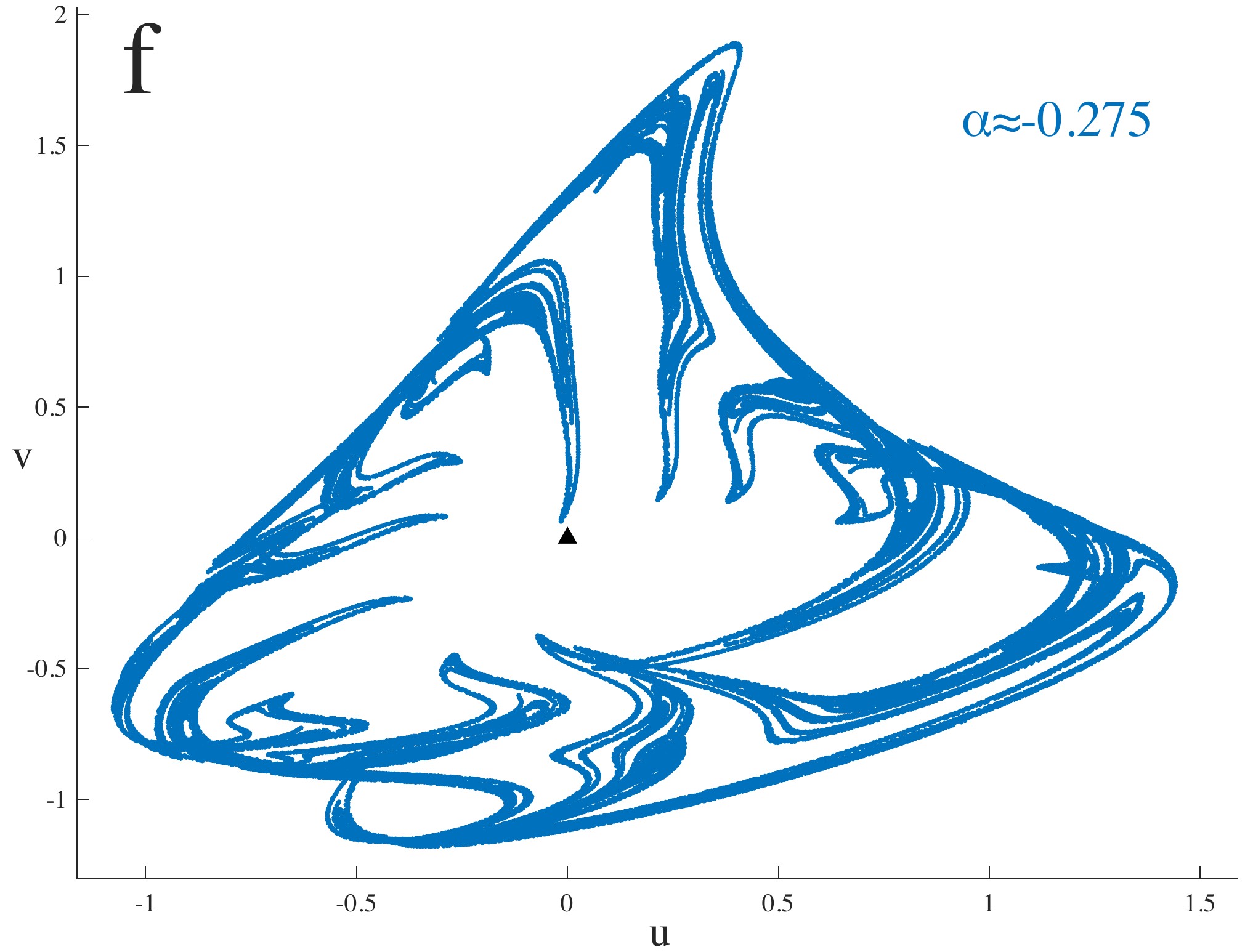}{Attracting orbits along \Eq{SilverSegment}, a segment in \Fig{LorenzTongues}, projected onto the plane \Eq{UVPlane}. The black triangles are the fixed point, $\xi_-$. Each panel contains orbits at the indicated values of $\alpha$ in the same color as the orbit. The panels correspond to the intervals labeled in \Fig{IrrationalInfo}(right).
(a) Four invariant circles near $\alpha_{s0}$. 
(b) Period-24 (blue), period-31 (yellow) and an invariant circle (green). 
(c) Period-608 (blue), chaotic  with $\mu_T=0.14771$ (yellow), and period-62 (green) orbits. 
(d) Chaotic orbit with $\mu_T=0.23646$.
(e) Period-7 orbit (blue) that doubles to period-14 (yellow and green). 
(f) Chaotic orbit with $\mu_T=0.48808$.  This attractor is shown in 3D in \Fig{LLChaoticAttractors}(a). Orbits become divergent for $\alpha \leq \alpha_{s1}$.}{LorenzSilverOrbitDevelopment}{0.3}

\subsection{Chaotic Attractors}\label{sec:ChaoticAttractors}

As we have known since Feigenbaum's classic 
studies of 1D maps, chaos can arise from a self-similar accumulation of period-doubling bifurcations, i.e., a period-doubling cascade. \cite{Guckenheimer02, Kuznetsov04} As we have seen for \Eq{Q3DMap}, cascades from the fixed point $\xi_-$,  \Fig{HenonCascadeTongue}, or from higher period orbits, at the `ends' of the tongues,  \Fig{LorenzTongues}, do indeed lead to chaos.

When $\delta$ is small, the resulting chaotic attractors can resemble that of the 2D H\'enon map. For case~\ref{SmallDeltaCase}, where $\delta=0.05$, the heteroclinic orbits of \Fig{HeteroclinicOrbits} show the beginning of the development of a H\'enon-like attractor through a period-doubling cascade of the fixed point. Continuing beyond the endpoint of the segment in \Fig{HenonCascadeTongue} leads to the attractor is shown in \Fig{ChaoticAttractors}(a), with a nearly 2D, horseshoe-like shape.
Note that for $\delta=0$, the map \Eq{Q3DMap} is essentially 2D,\footnote
{
It becomes a semi-direct product of a linear map and a 2D quadratic map. \cite{Hampton22}
} 
and for $(a,c,\alpha,\sigma)=(1,0, -1.4,-0.3)$, its dynamics correspond to the classic \hen map. \cite{Henon76} Even when $\delta = 0.05$, these parameters are at the very edge of the chaotic region after the period-doubling cascade of the fixed point, close to an arm of the period-$5$ shrimp in \Fig{HenonCascadeTongue}.
The attractor for this case is shown in \Fig{ChaoticAttractors}(b).
 
\InsertFigThree{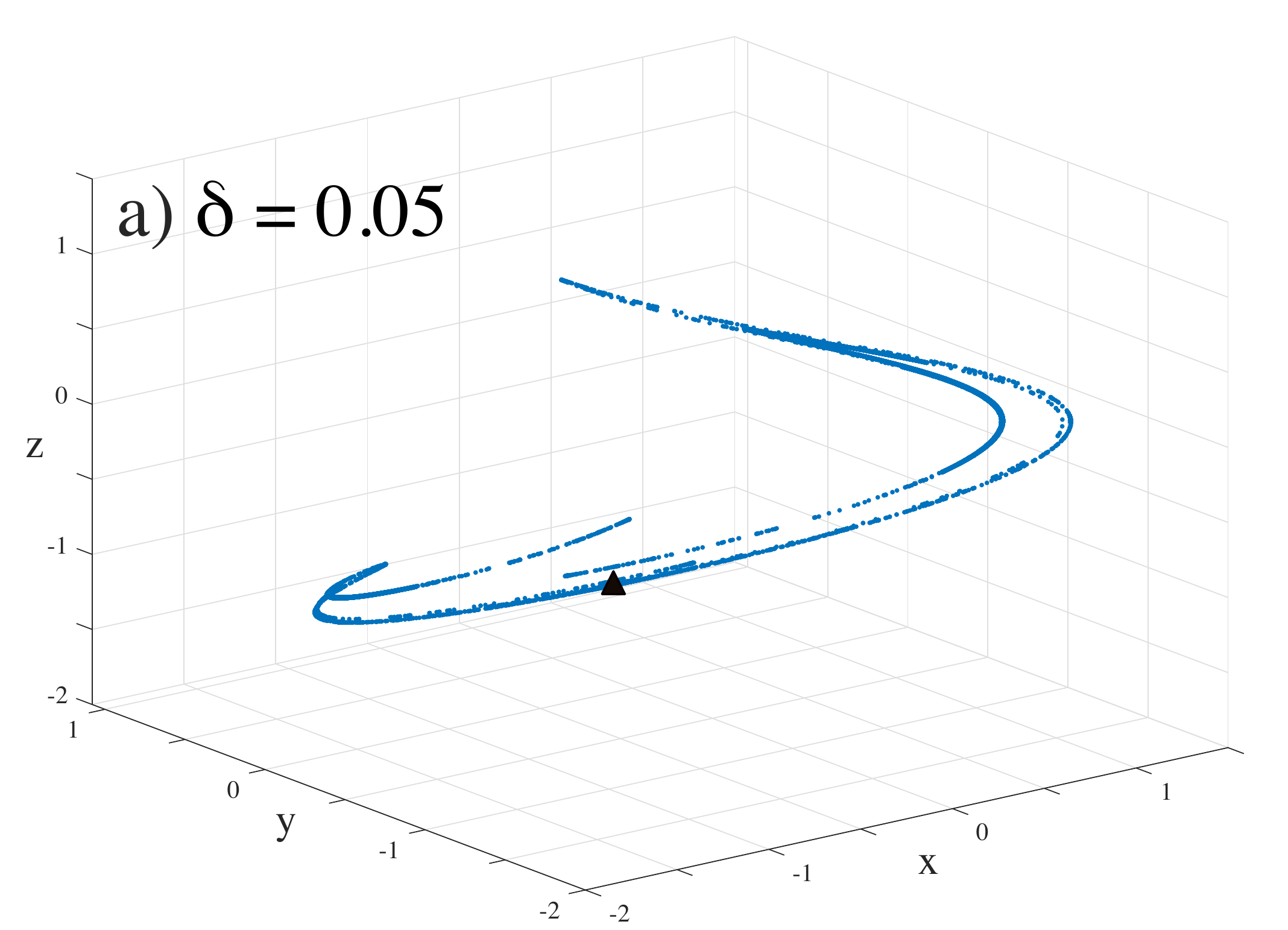}{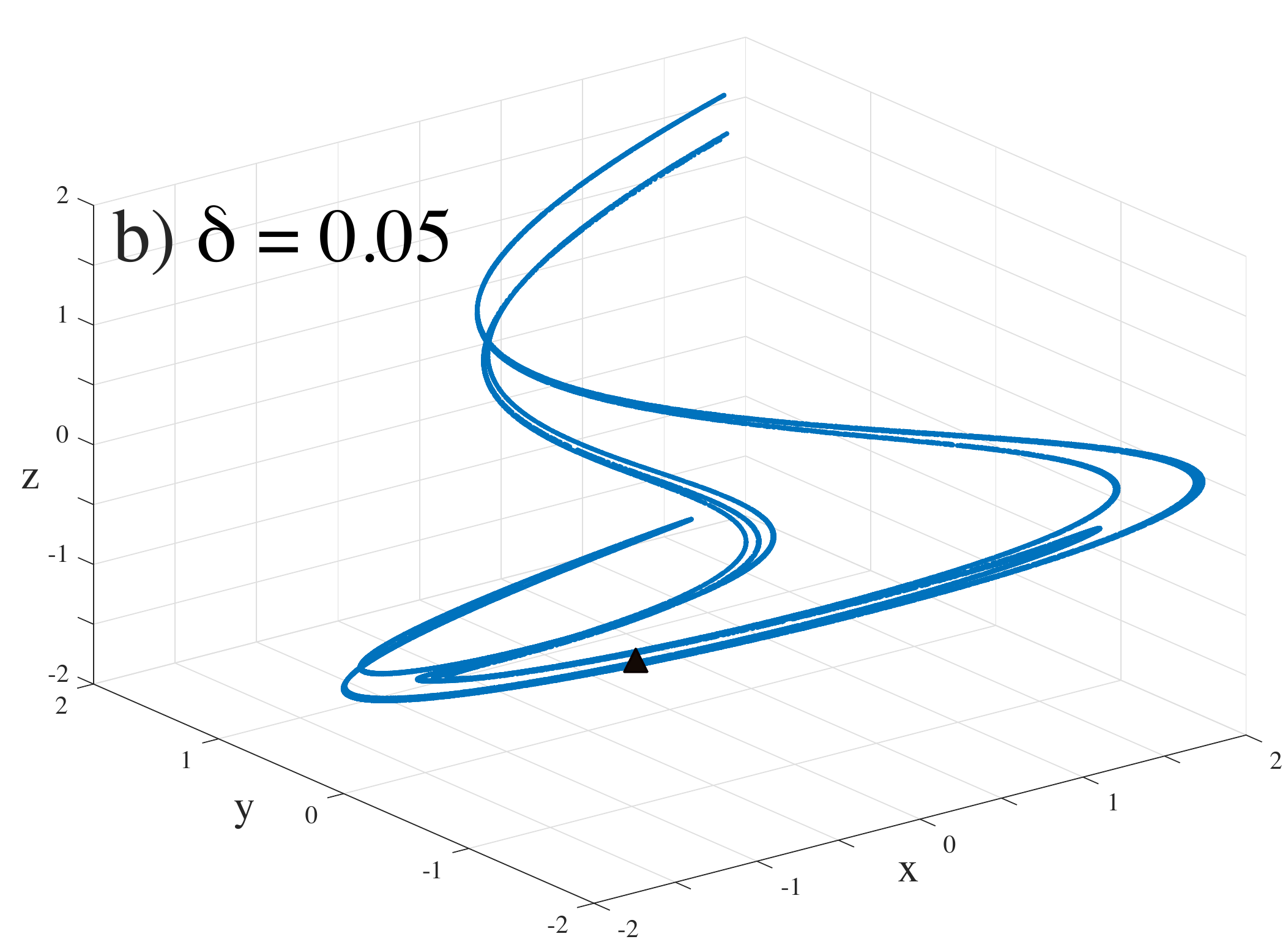}{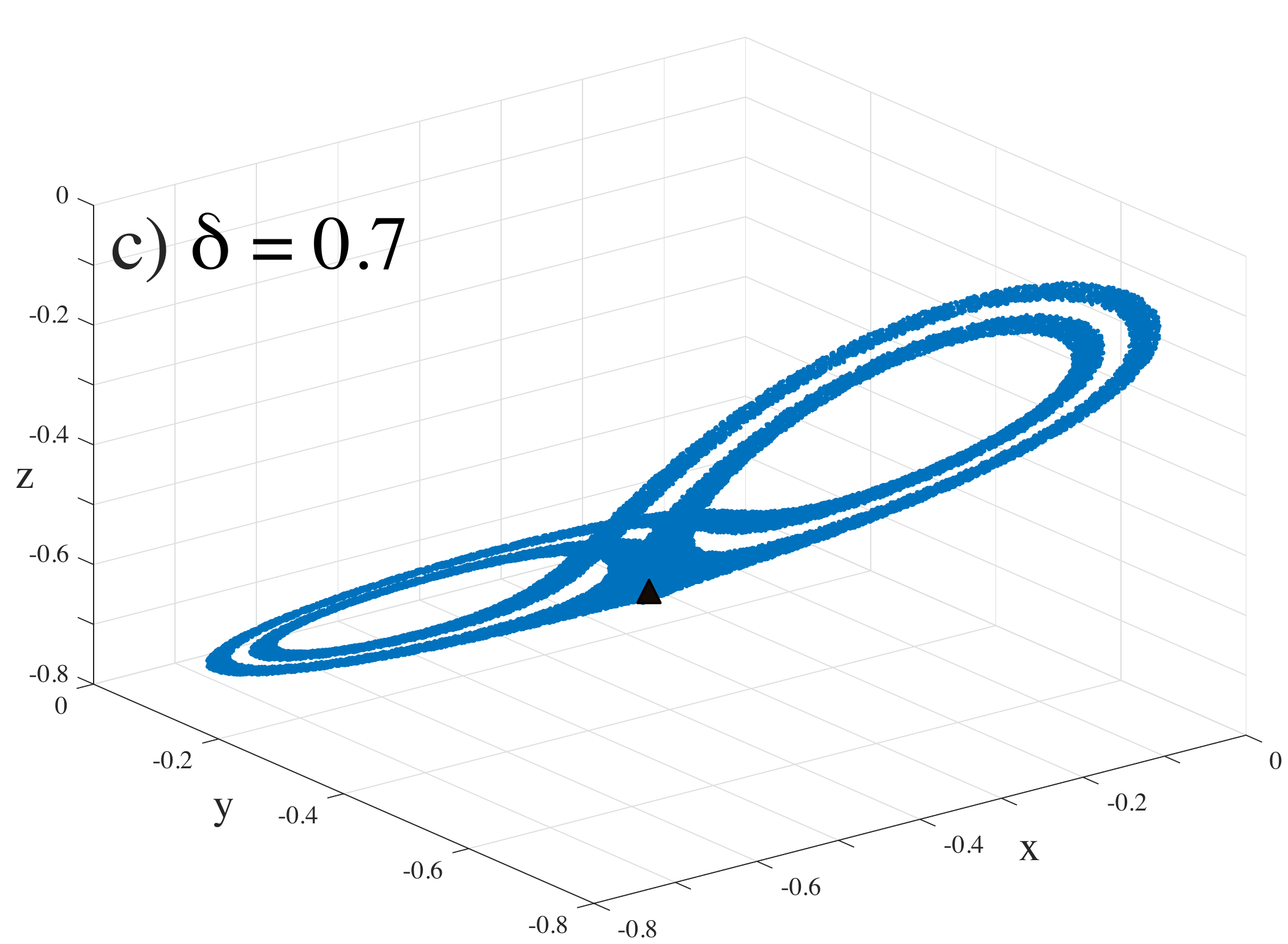}{Chaotic attractors illustrated as $500$ iterates 
of $100$ points in a ball of radius $0.01$ about the fixed point $\xi_-$ (black triangle). 
(a) Case~\ref{SmallDeltaCase} with $(\alpha,\sigma)\approx(-1.13599,-0.29130)$ following from the heteroclinic orbit seen in \Fig{HeteroclinicOrbits} after a period-doubling cascade. 
(b) Case~\ref{SmallDeltaCase} with $(\alpha,\sigma)=(-1.4,-0.3)$, near the parameters of the classic H\'enon attractor.
(c) A Lorenz-like attractor for case~\ref{LorenzLikeCase} with $(\alpha,\sigma)=(0,-0.815)$.
}{ChaoticAttractors}{0.3}

In \citeInline{Gonchenko21a}, parameters were found so that a 3D quadratic map conjugate to \Eq{Q3DMap} has a a discrete Lorenz-like attractor. The corresponding parameters for \Eq{Q3DMap} are $(a,c,\delta,\alpha,\sigma)=(1,0,0.7,0,-0.815)$. This case lies within the chaotic region below the NS bifurcation of the period-two orbit shown in \Fig{ResonantEx}(b). The resulting Lorenz-like attractor (with a lacuna) is shown in \Fig{ChaoticAttractors}(c). We refer to the works of Gonchenko et al. \cite{Gonchenko16,Gonchenko21a} for more discussion of such attractors.

In \Sec{Nonchaotic}, we saw the development of chaotic attractors as $\alpha$ decreased along the segment \Eq{SilverSegment}, recall \Fig{LorenzSilverOrbitDevelopment}(c,d,f). A three-dimensional plot of panel (f) is shown in \Fig{LLChaoticAttractors}(a) to better illustrate that it appears to lie near a paraboloid that opens up in the positive direction along the line $x=y=z$, which is near the local unstable manifold of the fixed point $\xi_-$.
To illustrate some of the variations in geometry that can occur, five additional cases, using parameters in the gray region of \Fig{ResonantEx}(b) for case~\ref{LorenzLikeCase}, are also pictured in \Fig{LLChaoticAttractors}. As before, each of these appears to lie near a paraboloid. The attractors in panels (a) and (b) have arms or tentacles, some of which appear to go towards the fixed point $\xi_-$. By contrast, in panels (b) and (c), the attractors more closely resemble invariant circles with additional folds. Finally, in panels (e) and (f) the attractors have an internal flower-like structure that fills out an annular region on the paraboloid.

\InsertFigSixV{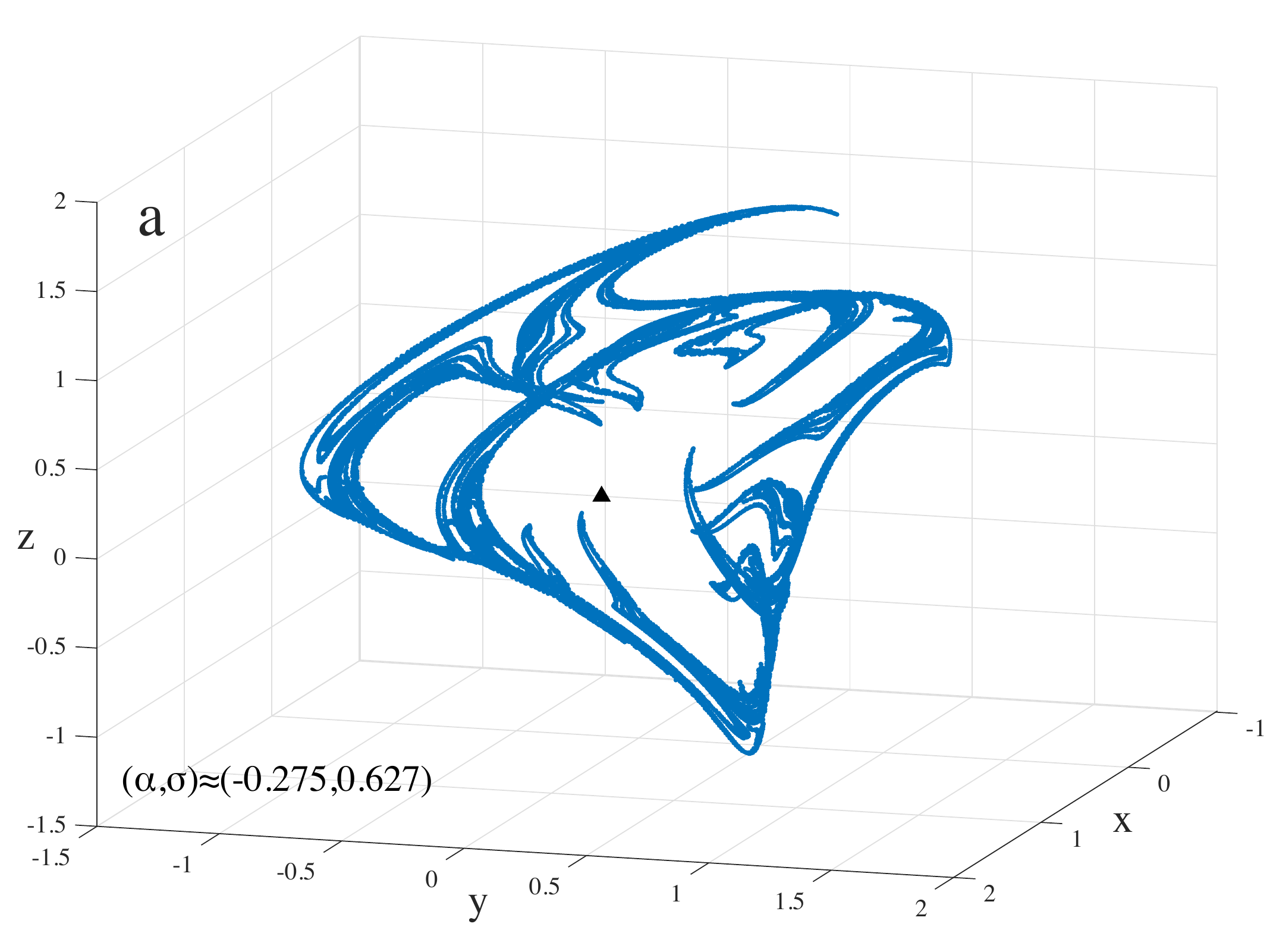}{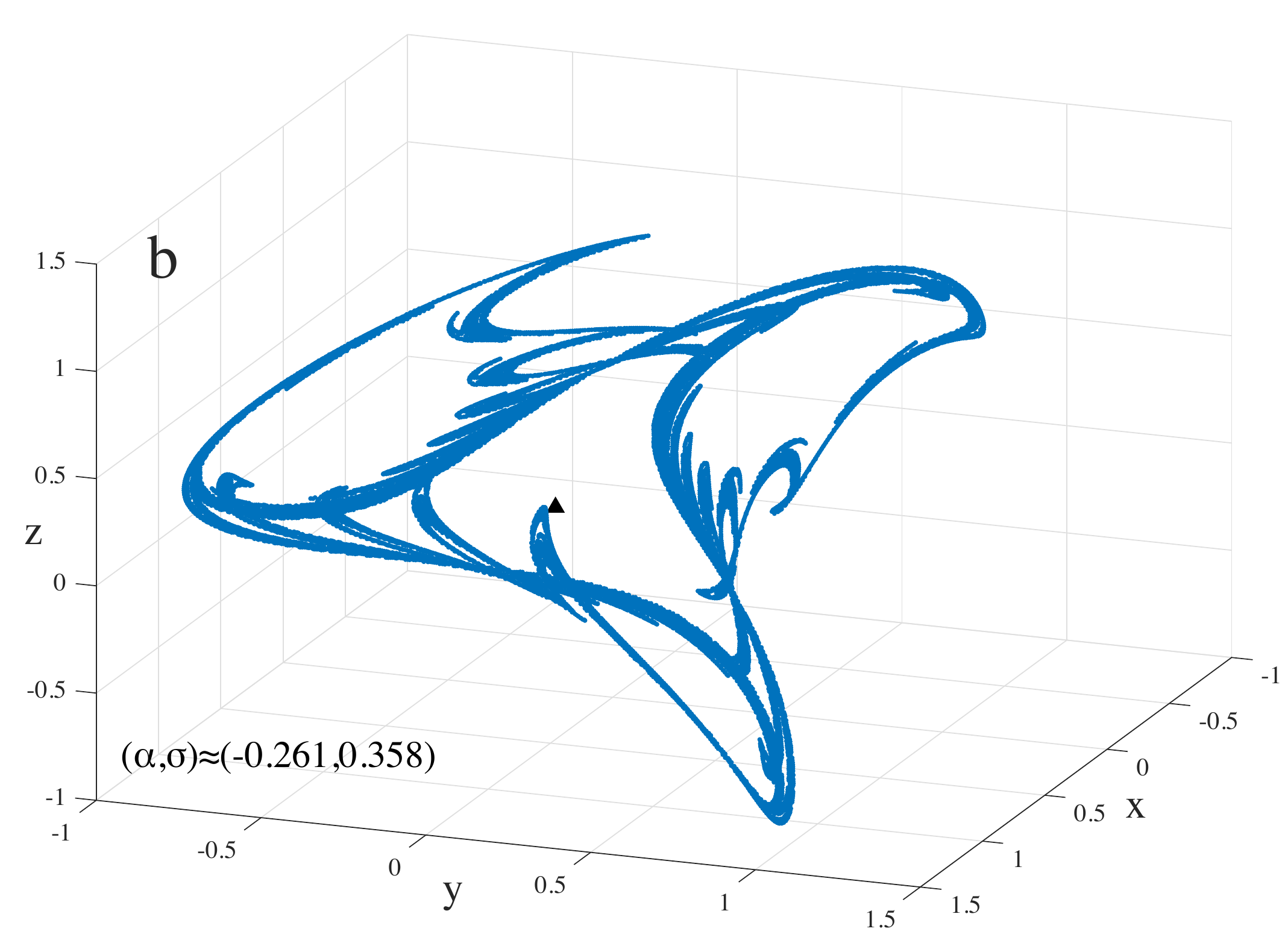}{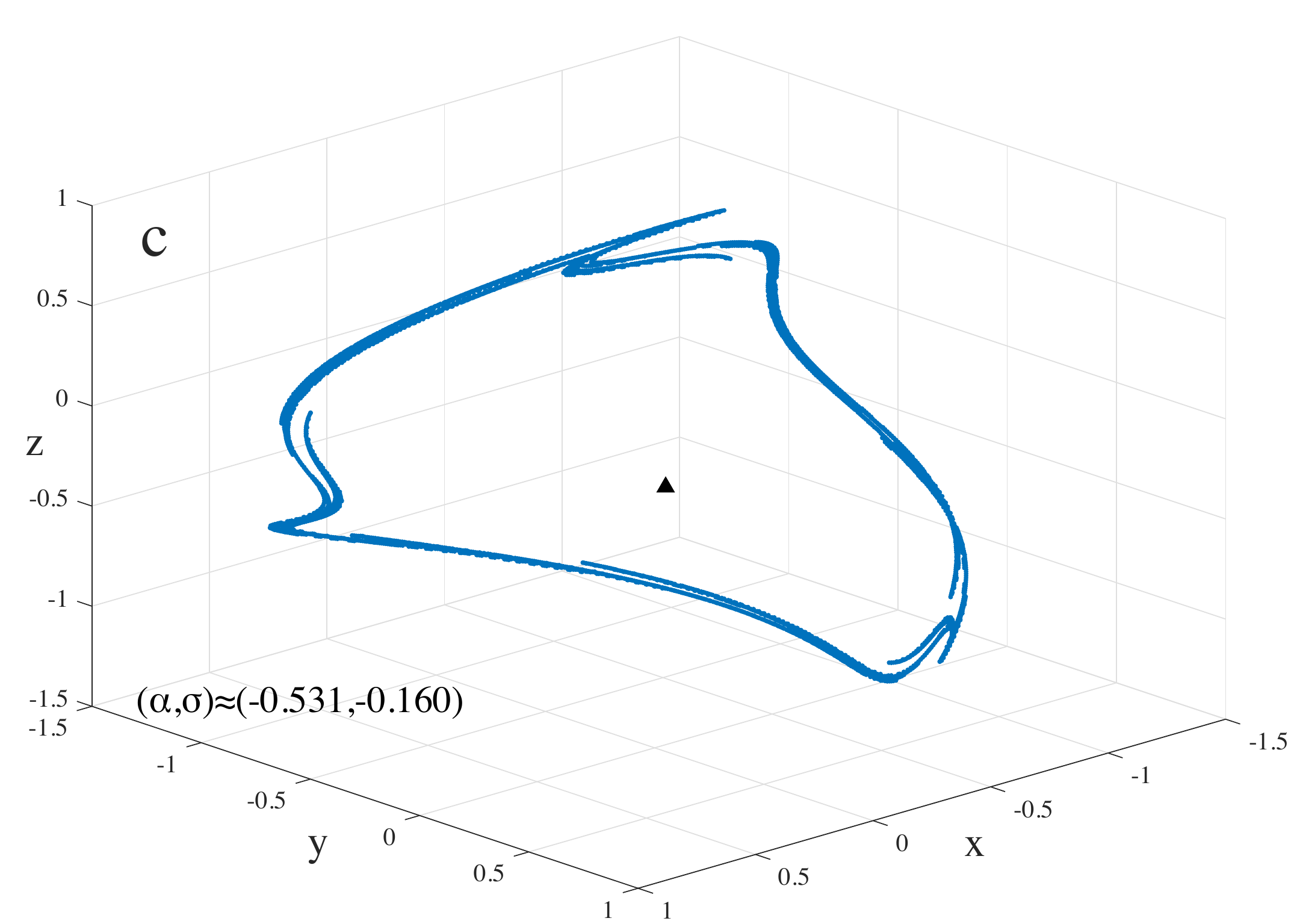}{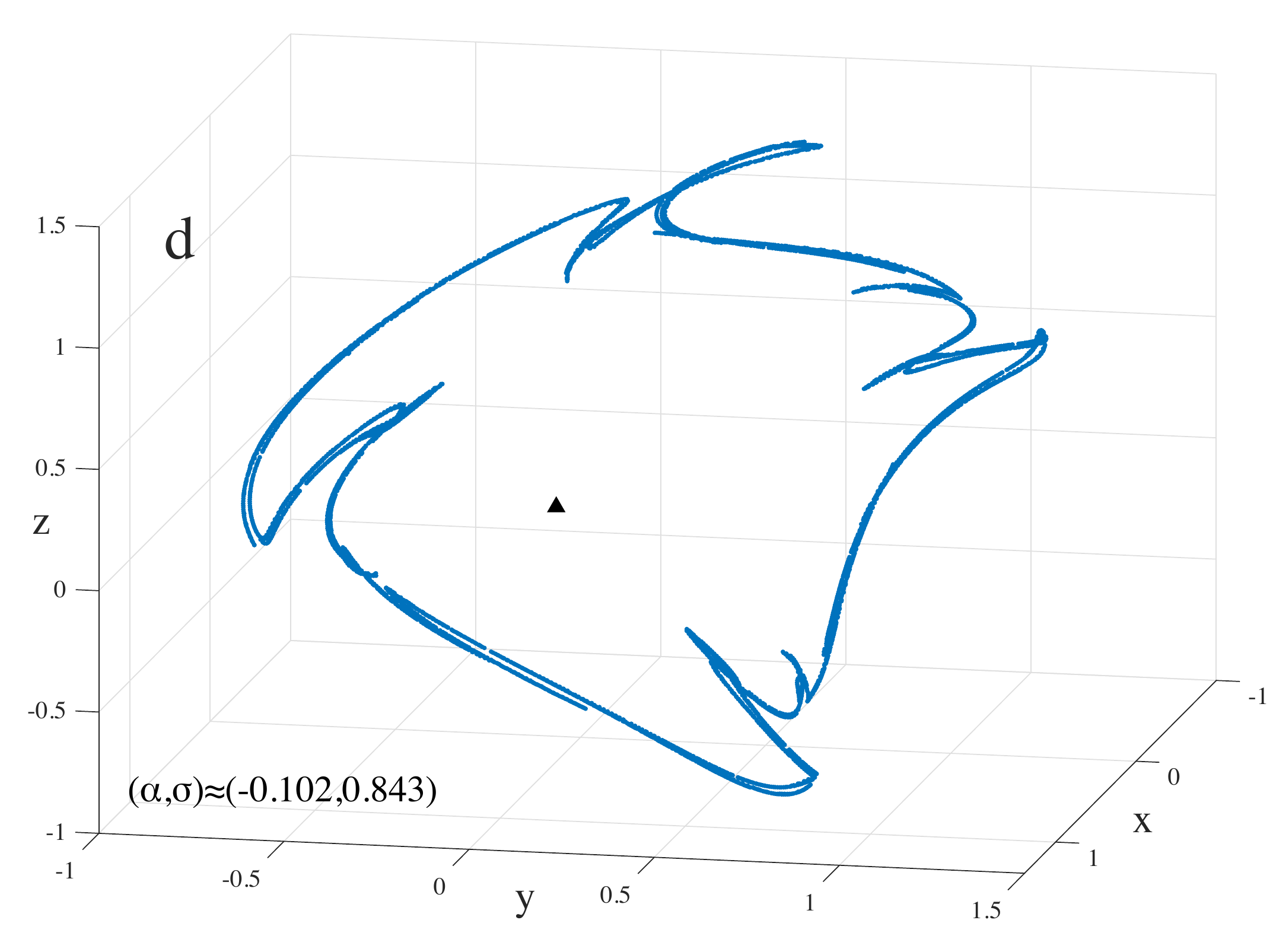}{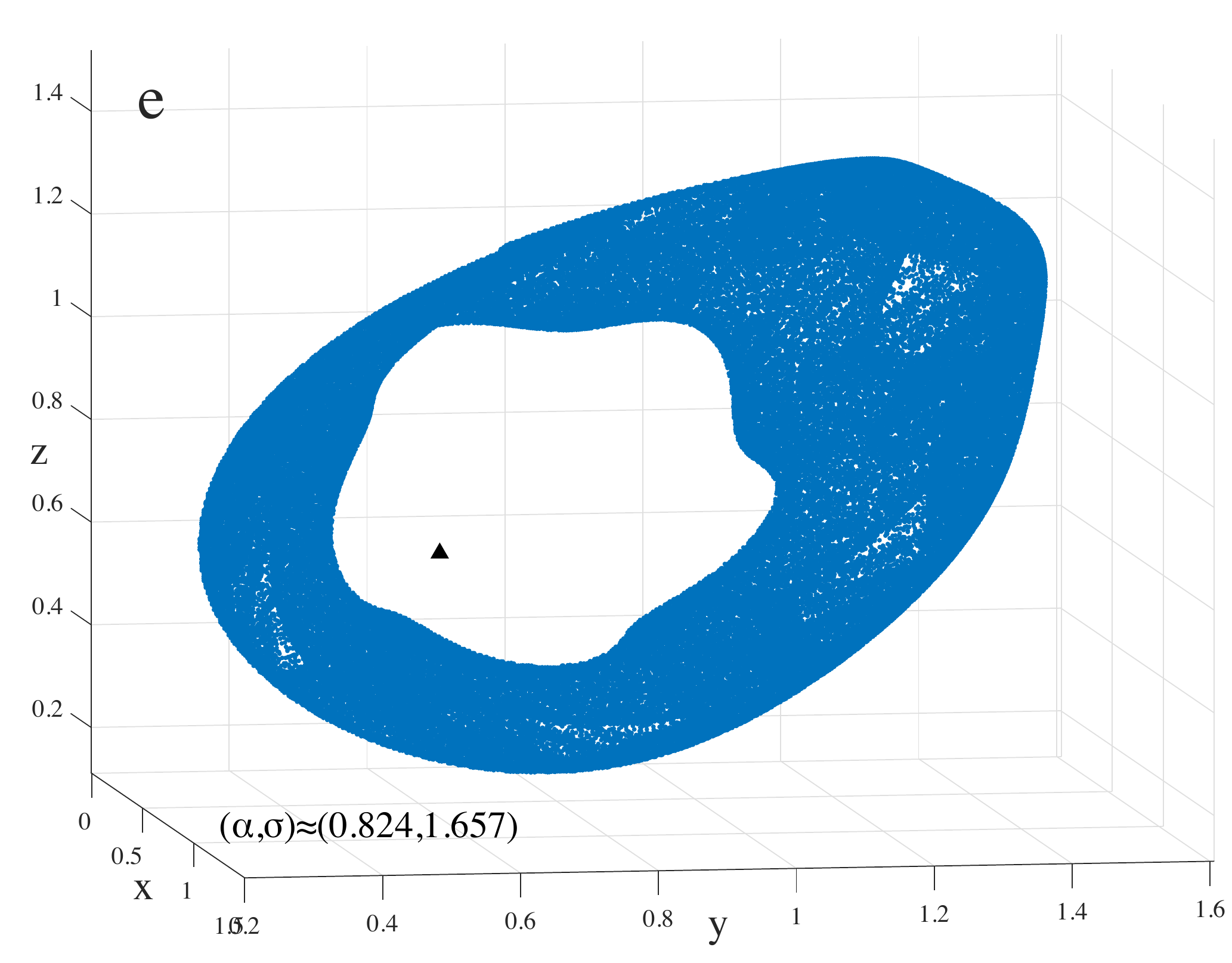}{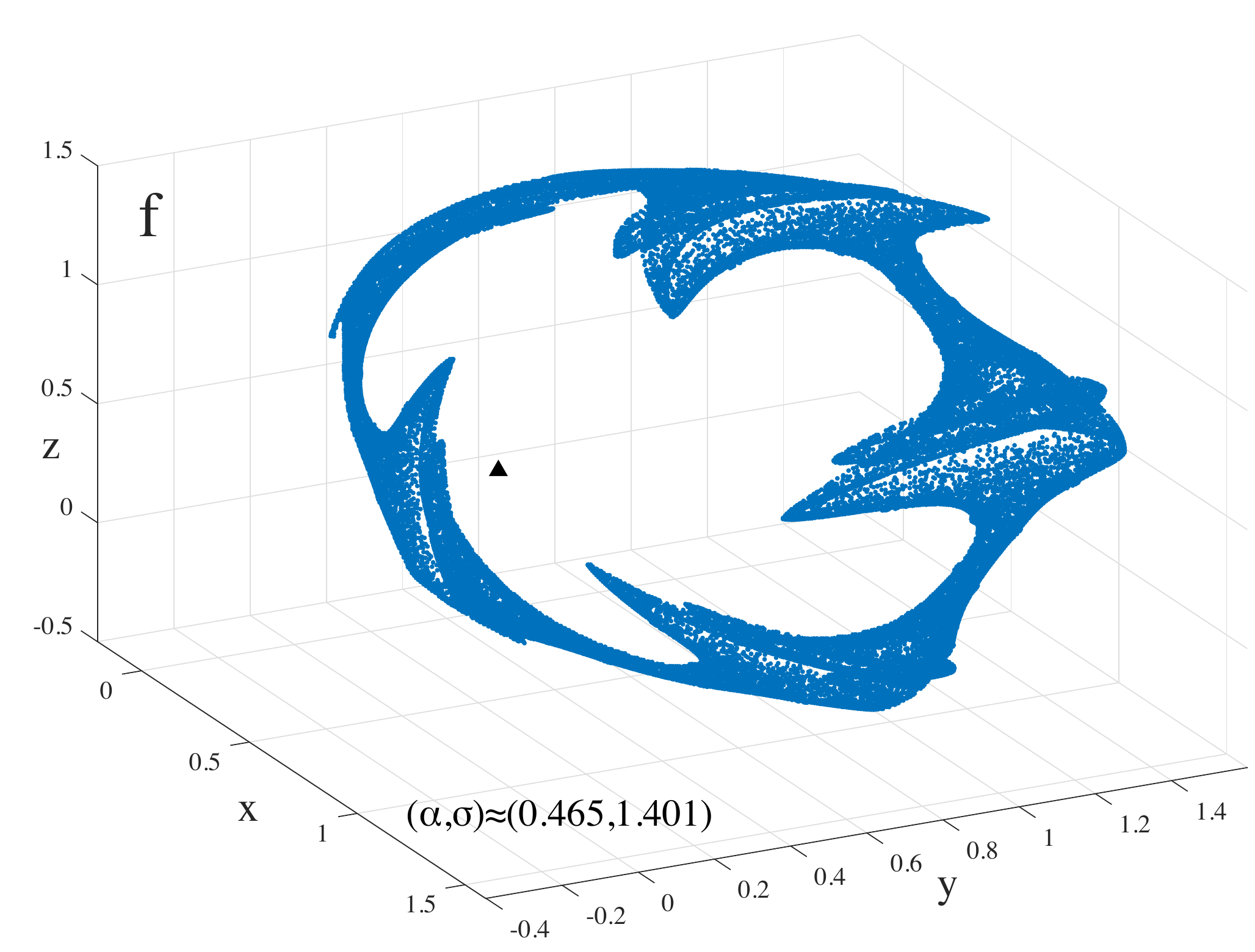}
{Chaotic attractors found using the method described in \Sec{Nonchaotic}: a point near $\xi_-$ (black triangle) is first iterated 5000 steps to remove transients, and then $T$ additional points ($T$ is either the return time \Eq{FirstReturn} or at most $5(10)^5$) are plotted.
(a) A 3D view of the attractor of \Fig{LorenzSilverOrbitDevelopment}(f) for case~\ref{LorenzLikeCase}. 
(b)-(f) Attractors for case~\ref{LorenzLikeCase} for five parameter cases (shown in each panel) taken from the chaotic (gray) regions of \Fig{ResonantEx}(b).
}{LLChaoticAttractors}{.45}

\section{Conclusions}

Previous research on quadratic 3D maps has focused on the volume-preserving case\cite{Lomeli98, Dullin09a} or on the existence and development of chaotic attractors. \cite{ Gonchenko05b,Gonchenko16,Gonchenko21a} Here, we have explored a broader range of parameters and studied periodic and aperiodic, regular and chaotic attractors. 

The simplest bifurcations of a 3D map were discussed in \Sec{GeneralBifTheory} using the trace and second trace of the Jacobian as primary parameters and fixing the Jacobian determinant. These results were applied to the fixed points of the map \Eq{Q3DMap} in \Sec{BifofFixedPts}. We focused on two primary parameters: the more ``structural'' parameter $\sigma$, that controls the \textit{type} of bifurcation and, what can be viewed as the ``primary unfolding'' parameter, $\alpha$. 
Note that in our previous work on  anti-integrability, it was the limit $\alpha \to -\infty$ that corresponded to a non-deterministic limit where the dynamics is conjugate to a shift on a set of symbols.\cite{Hampton22}

For our numerical studies, we chose two cases for the Jacobian $\delta$: a strongly contracting case~\ref{SmallDeltaCase}---where the map is nearly 2D---and a moderately contracting case~\ref{LorenzLikeCase}.

We showed in \Sec{BddOrbits} that all bounded orbits of the orientation preserving 3D \hen map lie within a cube about origin as illustrated in \Fig{BddVolumes}. Most of the region of bounded orbits correspond to  nonchaotic situations, which could be an attracting fixed point or orbits that arise from this point by doubling or Neimark-Sacker bifurcations. However this figure also shows protruding spikes from the region of stability, that may be
related to attractors not born from the fixed point. We hope to study these further in the future.

To classify the behavior of bounded orbits, we computed resonant regions in parameter space in \Sec{PeriodicAttractors}; these are analogous to the Arnold tongues of circle maps. The resulting partition of the bounded region, \Fig{ResonantEx},  shows periodic and aperiodic attractors. Using this, we are able to understand the development of attracting periodic orbits and visualize their codimension-one and -two bifurcations. Note that our computations used the attractor arising from a single initial condition. There will be cases with multiple attractors and cases for which the chosen orbit is unbounded even when there might be attractors elsewhere. But given the similarity between \Fig{BddVolumes} and \Fig{ResonantEx}, the latter possibility is rare. We plan to investigate the more complex, outlying cases in future research.

Aperiodic attractors (defined to be attractors with period greater than $90$), were studied in \Sec{NonresonantAttractors}. We also followed the evolution of invariant circles along several curves in parameter space
starting at points on the NS bifurcation curve where the rotation number was irrational.
We did not attempt to follow a circle with fixed rotation number,
though we expect such a curve exists in a two-dimensional parameter space and plan to do this in future research.
Along the parameter curves that we did follow, the invariant circle has
a varying rotation number. When this becomes rational, the circle 
undergoes a resonant bifurcation, generically breaking up into a pair of periodic orbits.
As the curve leaves a resonant tongue, an invariant circle can reform. In some cases
the destruction of the circle gave rise to chaotic attractors.


Some of the chaotic attractors we studied are well-known: the discrete Lorenz and \hen attractors found in \citesInline{Gonchenko21a,Henon76,Hampton22}. 
More unusual are the chaotic attractors with a paraboloid structure that arise from the destruction of the invariant circles, recall \Fig{LLChaoticAttractors}. These are not classified in \citeInline{Gonchenko16} and are not obviously related to the 3D generalized horseshoes studied by \citeInline{Zhang16}. We believe that further study of such cases could lead to a broader understanding of attractors that lie within higher-dimensional generalizations of the Smale horseshoe. We plan to analyze these further in future research.

\appendix
\section{Parameter space conversion}\label{app:ParameterConversion}

Here, the conditions for codimension-one and -two bifurcations as seen in the last column of \Tbl{Bifurcations} 
in terms of the trace and second trace are converted to conditions on $(\alpha, \sigma)$. We are interested in the fixed point $\xi_-$ of \Eq{Q3DMap}, given by \Eq{FixedPoints}, thus, $t$ and $s$ are given by \Eq{tsFixedPts}, and $d = \delta$.
We assume that $\tau = 0$, as in \Eq{ParamSpace}.

Provided that there are no singularities, the codimension-one bifurcation curves are easily found using \Eq{alphaBif}:
\bsplit{alphasigma}
&\text{(SN)}  &\alpha  &=\tfrac{1}{4}(\sigma-\delta+1)^2, \\
&\text{(PD)}  &\alpha  &=x_{PD}(2x_{SN}-x_{PD}),\\
&\text{(NS)}  &\alpha  &=x_{NS}(2x_{SN}-x_{NS}),
\esplit
for $x_{SN}=\tfrac{1}{2}(\sigma-\delta+1)$, and $x_{PD}, x_{NS}$ given by, \Eq{x_pd}, \Eq{x_NS}, respectively.
The NS bifurcation is restricted to the interval $\delta-2 < (2a+b)x_{NS}<\delta+2$.

Again, provided there are no singularities, the codimension-two bifurcations are all points that satisfy \Eq{alphaBif}, but also require an expression for $\sigma$. The SNf bifurcation occurs at $(t,s)=(-\delta,-1)$. Using $t$ to solve for the fixed point, then gives
\begin{align*}
  &\text{(SNf)}&  x_{SNf} &=-\frac{\delta}{2a+b} , \\
  & & \alpha &=x_{SNf}(2x_{SN}-x_{SNf}), \\
  & & \sigma &=-1+(b+2c)x_{SNf},
\end{align*}
where $x_{SN}$ is dependent on $\sigma$.

A Neimark-Sacker bifurcation with rotation number $\omega$ occurs at $(t,s)=(2\cos{(2\pi\omega)}+\delta,\delta(t-\delta)+1)$. Solving for the fixed point then gives
\begin{align*}
&\text{(R$\omega$)} &    x_{R\omega} &=\frac{2\cos{(2\pi\omega)}+\delta}{2a+b},  \\ 
& &    					\alpha &=x_{R\omega}(2x_{SN}-x_{R\omega}),\\
& &   					\sigma &=2\delta\cos{(2\pi\omega)}+1+(b+2c)x_{R\omega},
\end{align*}
where, again, $x_{SN}$ is dependent on $\sigma$.

Lastly, recall that double multipliers, $\lambda_{1,2}=r\in\bR$, occur on the parametric curves \Eq{DoubleEval}.
Using the same process as above yields, 
\begin{align*}
  &\text{($\lambda_1=\lambda_2$)} &  x_r &=\frac{\delta+2r^3}{r^2(2a+b)}, \\
  & &							    \alpha &=x_r(2x_{SN}-x_r),\\
  & &								\sigma &=2\tfrac{\delta}{r}+r^2+(b+2c)x_r,
\end{align*}
for $x_{SN}$ dependent on $\sigma$. Since we are only concerned with $x_-$ and we know $x_-<x_{SN}$, we choose $r$ such that $x_r<x_{SN}$. When enforcing \Eq{ParamSpace}, we obtain the range $|r| > \sqrt{\delta}$, otherwise $x_r=x_+$.

\begin{acknowledgments}
The authors acknowledge support from the National Science Foundation grant DMS-1812481.
\end{acknowledgments}

\section*{Data Availability Statement}
The data that supports the findings of this study are available within the article.  

\bibliography{AIBibliography}

\end{document}